# Determining optimal thermal energy storage charging temperature for cooling using integrated building and coil modeling

Ju-Hong OH[1], Seon-In Kim[2], Eui-Jong KIM [2,3]*

[1] Industrial Science and Technology Research Institute, Inha University, Incheon 22212, Korea; jhoh@inha.ac.kr

[2] Department of Smart City Engineering, Inha University, Incheon 22212, Korea; seonin97@inha.edu, ejkim@inha.ac.kr

[3] Department of Architectural Engineering, INHA University, Inha-ro 100, Michuhol-gu, Incheon, 22212, South Korea; ejkim@inha.ac.kr

**\*** Correspondence: ejkim@inha.ac.kr; Tel.: +82-32-860-7589

**Abstract**

Thermal energy storage (TES) systems coupled with heat pumps offer significant potential for improving building energy efficiency by shifting electricity demand to off-peak hours. However, conventional operating strategies maintain conservatively low chilled water temperatures throughout the cooling season, a practice that results in suboptimal heat pump performance. This study proposes a physics-based integrated simulation framework to determine the maximum feasible chilled water supply temperature while ensuring cooling stability. The framework integrates four submodels: relative humidity prediction, dynamic cooling load estimation, cooling coil performance prediction, and TES discharge temperature prediction. Validation against measured data from an office building demonstrates reliable accuracy across all sub-models (e.g., CVRMSE of 9.3% for cooling load and R² of 0.91 for peak-time discharge temperature). The integrated simulation reveals that the proposed framework can increase the daily initial TES charging temperature by an average of 2.55 °C compared to conventional fixed-temperature operation, enabling the heat pump to operate at a higher coefficient of performance. This study contributes a practical methodology for optimizing TES charging temperatures in building heating, ventilation, and air conditioning (HVAC) systems while maintaining indoor setpoint temperatures.

# 1. Introduction

Approximately 30% of the global final energy consumption occurs in the building sector, with heating, ventilation, and air conditioning (HVAC) systems accounting for 40–60% of this consumption [1,2]. Consequently, improving the energy efficiency of buildings has been identified as a critical priority by major organizations, including the International Energy Agency (IEA) and ASHRAE, to achieve carbon neutrality [3,4]. In this context, within the broader effort to reduce building-sector energy consumption, heat pump systems integrated with thermal energy storage (TES) have emerged as a promising solution for enhancing grid stability and reducing operational costs by shifting electricity demand to off-peak nighttime hours [5,6].

From a thermodynamic perspective, the coefficient of performance (COP) of a heat pump is inversely proportional to the difference between the condensation and evaporation temperatures. Therefore, setting the chilled water production temperature as high as possible during summer cold storage operations is advantageous for system efficiency [7,8]. Previous studies have reported that the COP improves by approximately 2–4% for every 1 °C increase in the chilled water supply temperature [9]. Notably, Wei et al. [10] demonstrated a 14.1% efficiency improvement through supply temperature optimization in a ground-source heat pump system. However, in practice, conservative operating strategies that maintain low fixed setpoints based on design criteria are conventionally applied to account for the uncertainties in daytime peak loads [11]. Such conservative approaches cause the system to produce chilled water at unnecessarily low temperatures, even under partial-load conditions or during intermediate seasons, thereby lowering the evaporation temperature of the heat pump and ultimately degrading the overall system efficiency.

Previous research on improving the energy efficiency and reducing costs in TES-coupled heat pump systems can be broadly categorized into two main areas: macro-scheduling studies and micro-dynamic control studies of subsystems, depending on the optimization targets and modeling fidelity.

Early studies primarily focused on macro-level scheduling to manage the overall energy flow of systems. These studies emphasized the optimization of system-level variables such as electricity tariffs and equipment capacities, rather than the detailed thermal behavior of buildings [12–21]. Henze et al. [12] compared priority control strategies for chillers and TES to analyze the operational costs, whereas Sebzali et al. [13,14] optimized the heat source equipment capacity and electricity consumption through energy time-shift strategies that relocate daytime peak loads to nighttime. Chen et al. [15] derived optimal design alternatives from a life-cycle cost (LCC) perspective using dynamic programming. However, these macro-level scheduling studies assumed the building loads to be predetermined fixed values or used simplified models, thereby failing to precisely reflect real-time variations in indoor environments and the physical constraints of the equipment.

To address these limitations, dynamic control and modeling techniques at the subsystem level (building and air handling unit) have been actively investigated to enhance the control precision of TES systems [22–30]. Although many of these subsystem-level studies originated from general HVAC systems without TES, their findings are essential for improving the load prediction accuracy and heat exchange efficiency, which are critical for TES operations. Chinde et al. [27] demonstrated that optimal control of HVAC systems can be achieved using only input–output data without system identification processes through data-enabled predictive control (DeePC). Choi et al. [28] applied model predictive control (MPC) based on resistance-capacitance (RC) gray-box models to heat pump systems, verifying energy savings of approximately 4.96–9% and cost reductions of 15.5–20.4%. Additionally, for an accurate analysis of the physical components, Oh et al. [29] proposed a model calibration method using air handling unit (AHU) operational data, and Jin et al. [30] developed a dynamic model of cooling coils to characterize the heat exchange performance under varying flow rates.

However, previous studies have addressed these elemental technologies individually and have not integrated them to optimize chilled water supply temperatures in TES systems. Most studies either fixed the chilled water temperature at the design reference values or approached the problem solely from a heat pump COP perspective, failing to comprehensively consider questions such as whether the system can handle loads when the logarithmic mean temperature difference (LMTD) decreases owing to elevated chilled water temperatures, that is, when the heat exchange performance of the coil is degraded. Thus, coupling coil performance models with load models is essential for predicting the risk of indoor temperature control failure resulting from increased chilled water supply temperatures.

This limitation arises because increasing the chilled water supply temperature introduces an inherent trade-off: while a higher temperature improves the COP of the heat pump, it simultaneously causes a decline in the dehumidification and cooling capacity of the AHU coil. In particular, in cold storage systems utilizing off-peak electricity, the upper limit of the chilled water temperature that can stably handle indoor loads is not fixed. Instead, it must be treated as a dynamic constraint that varies moment by moment depending on the next-day building load and the thermodynamic state of the fluid inside the coil. In other words, beyond achieving simple energy savings, an integrated decision-making framework that systematically considers precise load prediction, the physical performance limits of heat exchangers, and the discharge capacity of TES is required.

Accordingly, this study proposes a physics-based integrated simulation framework capable of maximizing the chilled water supply temperature while ensuring cooling stability. The proposed framework was constructed by organically linking a relative humidity prediction model, a dynamic load model, a coil performance prediction model, and a TES model. Through this integrated modeling approach, the system calculates the maximum allowable chilled water temperature, which varies according to the real-time load and operating conditions. This approach enables the identification of the physical operational limit specifically, how high the chilled water supply temperature can be raised on the previous day without compromising cooling comfort and allows evaluation of the feasibility of the efficiency improvement potential by tracking the TES charging temperature relative to conventional fixed-temperature storage methods.

Summarizing the above, the research questions of this study are as follows:

1. Modeling and definition: Under variable cooling load conditions and coil capacity constraints, can a physical upper limit of the chilled water supply temperature be defined is sufficient to meet the cooling demand?

2. Operational Margin and Potential Assessment: When the derived dynamic temperature limit is applied, what margin exists for increasing the chilled water temperature compared with conventional fixed-temperature storage methods, and can an optimal TES charging temperature be identified?

The remainder of this paper is organized as follows. The problem statement is presented in Section 2. Section 3 describes the proposed modeling methodology, including relative humidity prediction, cooling load modeling, cooling coil modeling, and thermal energy storage modeling. Section 4 provides information on the case study, specifically the target building and system configuration. Section 5 presents simulation results and performance validation. Finally, Section 6 concludes the paper with a summary of the key findings and recommendations for future research.

## 2. Problem Statement

A thermal energy storage system coupled with a water source heat pump plays a crucial role in distributing electricity demand by producing (charging) chilled water at night using off-peak electricity and discharging it during the day, while simultaneously reducing operational costs through relatively inexpensive nighttime electricity rates. As mentioned in the Introduction, in addition to the benefit of lower electricity rates, efficient operation is hindered because chilled water is typically produced at a constant low setpoint temperature throughout the summer cooling season, primarily owing to difficulties in accurately predicting next-day loads and ensuring system stability.

To achieve energy savings by increasing the stored chilled water temperature in the TES, it is first necessary to determine whether the selected temperature satisfies daytime cooling load requirements. The heat exchange performance of the AHU cooling coil is determined by the temperature difference between the fluids (LMTD) and the mass flow rate. When the chilled water supply temperature is increased to improve the heat pump efficiency, the LMTD inside the coil decreases, implying a reduction in the heat exchange capacity per unit area of the coil. To compensate for this reduction, the chilled water flow rate must be increased; however, the physically available maximum flow rate and effective heat transfer area are limited. Therefore, a critical operating point exists at which even operation at the maximum allowable flow rate cannot offset the reduced heat exchange caused by the decreased LMTD, and the target cooling and dehumidification cannot be achieved, even when the chilled water temperature remains lower than the indoor temperature.

Furthermore, even for a TES operating at the same nominal set-point temperature, the upper limit of the chilled water temperature (maximum allowable chilled water temperature) required to satisfy the indoor set-point temperature is not fixed. Instead, this limit should be regarded as a time-varying parameter that fluctuates according to variations in the instantaneous load magnitude and cumulative daily thermal demand. However, conventional schedule-based control or static operating methods have structural limitations in that they fail to evaluate these dynamic load characteristics and the physical constraints of heat exchangers in a real-time, integrated manner. Owing to this uncertainty in load prediction and heat exchanger constraints, operators often apply excessive safety margins beyond what is strictly necessary to minimize the risk of cooling failure. These conservative margins ultimately act as a decisive factor in limiting the energy-saving potential of the system.

Fig. 1 schematically illustrates the relationship between the physical-limit temperature of the coil according to the cooling load conditions and the TES discharge characteristics, divided into (a) high-load conditions and (b) partial-load conditions. First, under the high-load scenario shown in Fig. 1(a), the limit inlet temperature of the coil is formed at a low level to handle the high cooling load. To ensure a stable cooling supply of the system at this time, the supply water temperature must not exceed the coil's limit temperature, not only at the peak load point, but also at the end of operation, when the TES water temperature rises to its highest level as the discharge progresses. Thus, managing the TES supply temperature trajectory such that it remains below the coil limit temperature throughout the operating period becomes a critical operational constraint.

In contrast, under the partial-load scenario in Fig. 1(b), the temperature limit that the coil can tolerate is maintained at a high level owing to the relatively low cooling demand. In this case, a sufficient thermal margin exists such that the temperature limit is not reached by the end of the operation, even as the TES water temperature increases. However, conventional control methods have the limitation of uniformly supplying low-temperature chilled water fixed to the design criteria without reflecting the real-time available range. Consequently, an unnecessarily excessive safety margin is generated between the maximum temperature that the coil can tolerate and the conservatively fixed supply temperature. This gap indicates that a substantial portion of the efficiency improvement potential associated with higher heat pump operating temperatures remains unutilized.

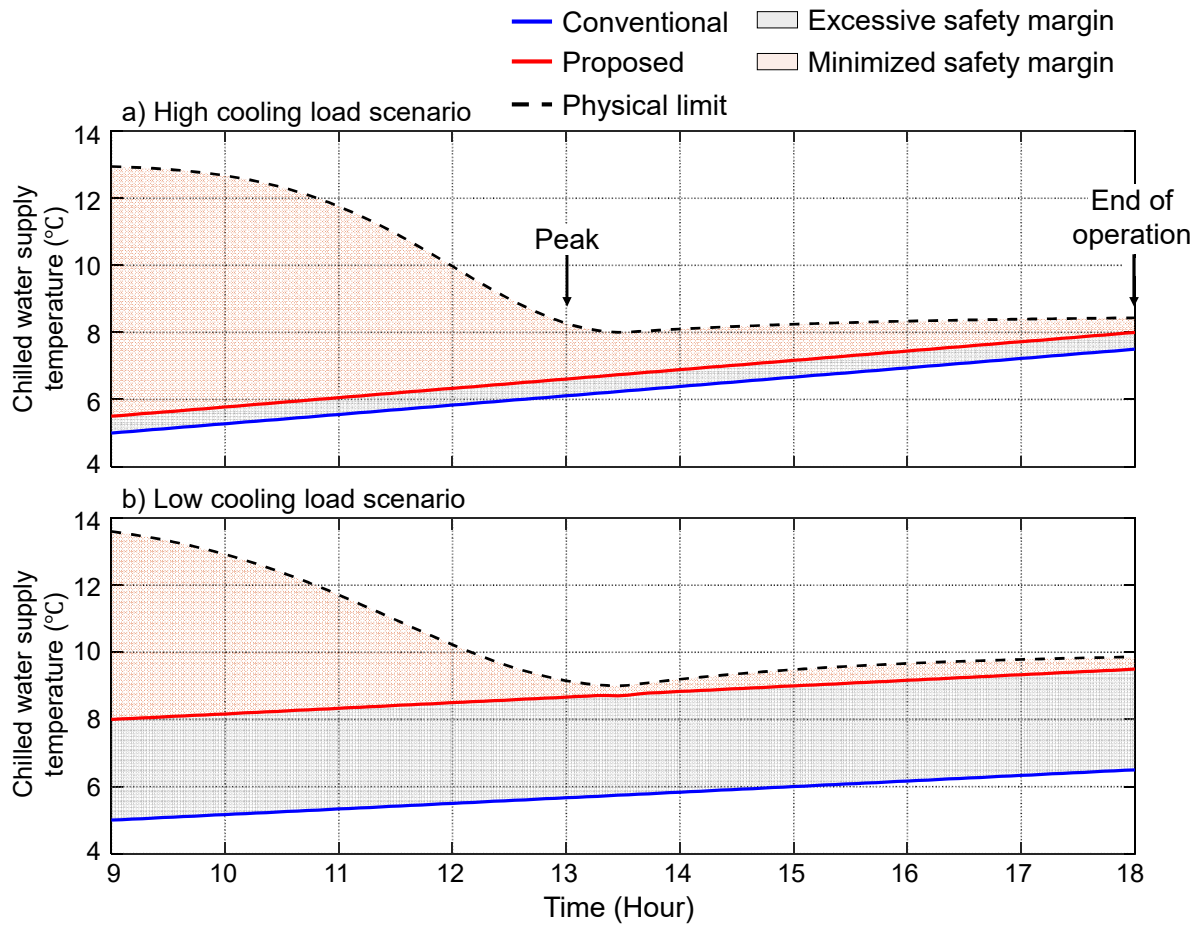

Fig. 1. Schematic comparison of chilled water supply temperature control strategies

Therefore, to improve the operational efficiency of heat pumps, an integrated decision-making framework that simultaneously considers building load prediction, physical performance limits of heat exchangers, and TES supply capacity is required to propose appropriate setpoint temperatures on a daily basis. Accordingly, this study constructs a physics-based integrated modeling framework that links these components and derives the optimal thermal storage temperature capable of ensuring cooling stability while maximizing overall system energy efficiency.

## 3. Method

### 3.1 Development process

Fig. 2 illustrates the integrated simulation framework proposed in this study for optimizing the TES charging temperature. The entire process consists of three stages, starting with data acquisition: demand prediction, system response simulation, and optimization and decision. In Stage 1, a zone load model was constructed based on the building envelope and system information, along with measured outdoor weather data. Using the load model, the indoor temperature and required cooling load were predicted based on operating and outdoor conditions. Additionally, the indoor humidity used for calculating the thermophysical properties, such as the enthalpy and specific heat of the coil inlet air, was predicted using a humidity prediction model.

Stage 2 involves predicting the TES discharge temperature ($T_{\text{TES,out}}$) and identifying the operational limits of the cooling coil. First, the next-day cooling load profile derived in Stage 1 was input into the TES discharge prediction model to predict the discharge temperature at the peak load point and at the end of the operation according to the initial charging temperature and cumulative load rate. On the water side, pipe heat losses were neglected, and the rated flow rate operation was assumed to simulate the coil inlet conditions. Although actual systems involve flow-rate control for stratification maintenance and partial load response, this study reflects this behavior through a simplified model based on an energy conservation law. Simultaneously, the air-side state variables of the AHU, including the indoor temperature and humidity, and the required cooling load derived in Stage 1 are input into the heat exchanger model. An inverse NTU-ε algorithm is then applied to back-calculate the maximum allowable chilled water inlet temperature ($T_{\text{Lim t}}$) to handle the given load under the given air conditions.

In the final stage, we determine whether the expected TES discharge temperature remains below the coil's allowable limit temperature ($T_{\text{TES,out}} \leq T_{\text{Lim t}}$) throughout the entire operating period. Owing to the characteristics of the TES, satisfying the conditions at the last moment of the daily schedule determines the suitability of the charging setpoint temperature. If the temperature feasibility conditions are not satisfied, the initial charging setpoint temperature is readjusted and the simulation is iteratively performed until an optimal operating point satisfying the conditions is determined.

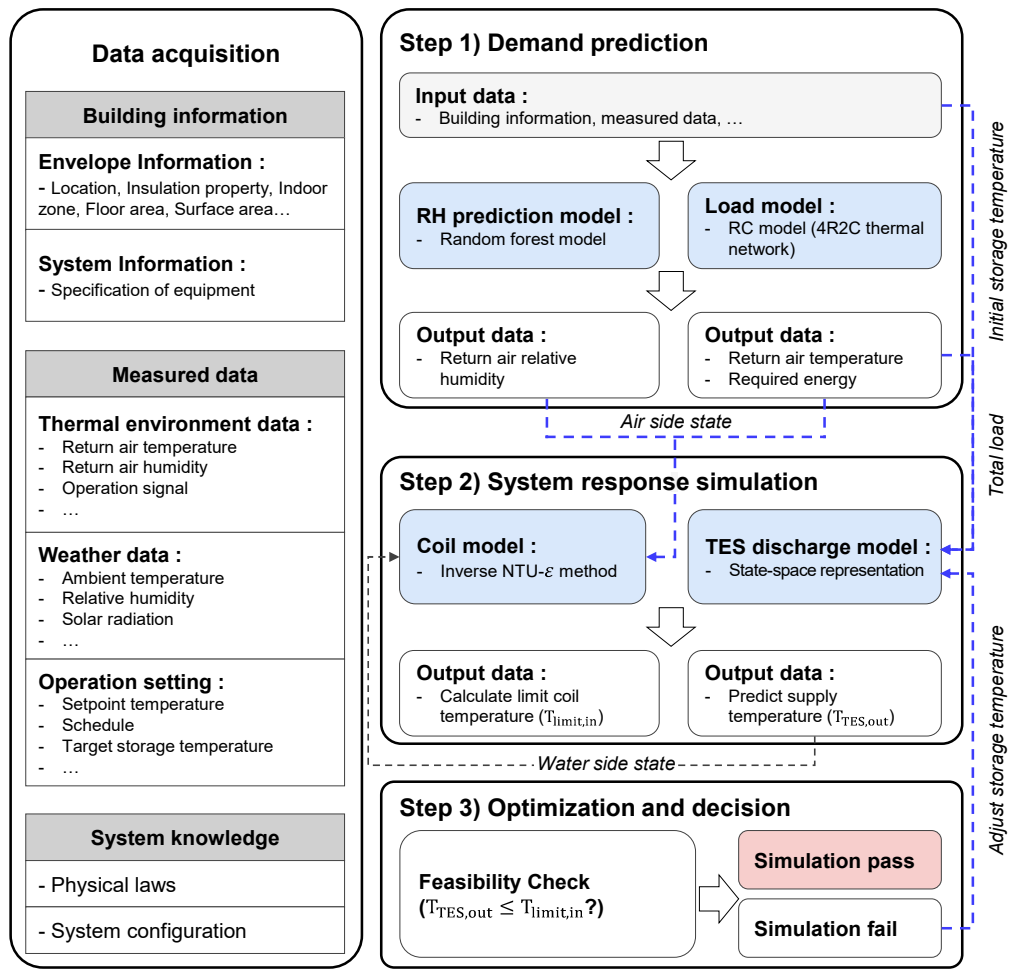

Fig. 2. Schematic diagram of the integrated simulation framework for optimizing TES discharge temperature

## 3.2 Relative humidity prediction modeling

In the analysis of the thermodynamic behavior of cooling systems, the indoor relative humidity is a key variable that determines air properties such as the enthalpy and specific heat of air. In particular, because the cooling process of moist air involves not only sensible heat but also latent heat changes owing to dehumidification, accurate humidity information is required for modeling the total heat transfer performance of the coil. However, indoor humidity is determined by complex and nonlinear factors, including infiltration, ventilation, occupant activity, and system operating status, making it difficult to predict it using simple physical equations or linear models. Therefore, in this study, a data-driven indoor humidity prediction model was developed using readily obtainable weather and AHU operating data.

To enhance the accuracy of the humidity prediction model, physics-informed feature engineering was performed, reflecting the thermodynamic relationships on a psychrometric chart, as shown in Fig. 3. The model input variables included not only dry-bulb temperature and relative humidity but also explicitly calculated values, such as dew point temperature and absolute humidity, representing the absolute moisture content in the air, and the vapor pressure ratio (ratio of outdoor vapor pressure to indoor saturation vapor pressure) related to the driving force for moisture transfer. The dew point temperature and absolute humidity were calculated using the Magnus–Tetens approximation [31, 32], as shown in Equation (1). Absolute humidity was calculated based on the saturation vapor pressure, as shown in Equation (2). where a and b are the saturation vapor pressure constants (17.27 and 237.7, respectively), e is the actual vapor pressure, and $e_s$ is the saturation vapor pressure, as given in Equations (3) and (4).

$$T_{\mathrm{dp}} = \frac{b \cdot (\frac{aT}{(b+T)} + \ln(RH/100))}{a - (\frac{aT}{(b+T)} + \ln(RH/100))} \tag{1}$$

$$\text{AH} = \frac{217 \times e}{T} \qquad (2)$$

$$e = (\text{RH}/100) \times e_s \qquad (3)$$

$$e_s = 6.12 \times \exp\left(\frac{17.67T}{T + 243.5}\right) \qquad (4)$$

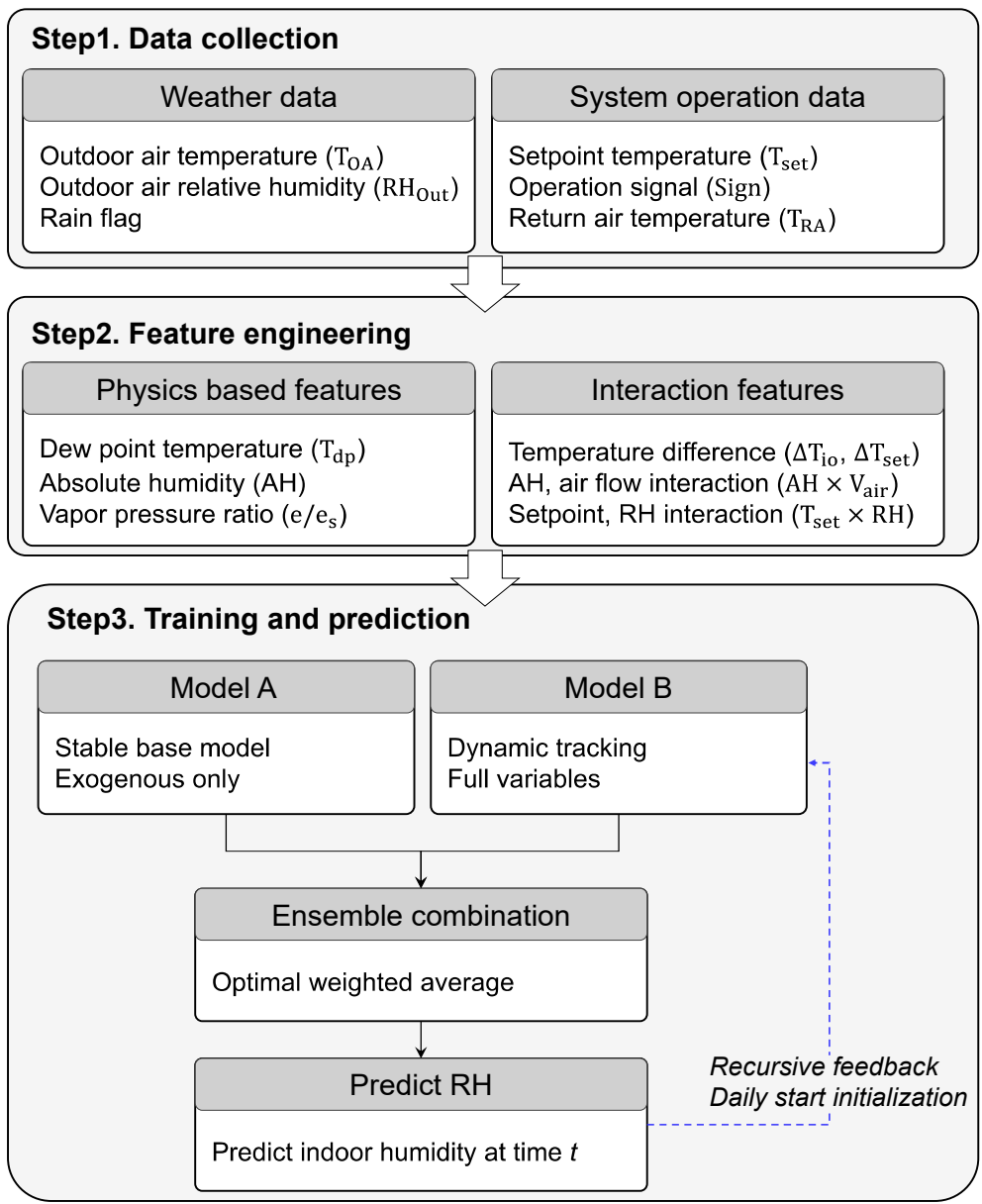

Fig. 3. Schematic diagram of the relative humidity prediction model using recursive feedback

Additionally, to capture the dynamic behavior of the system, derived variables representing operating states were introduced, such as the indoor–outdoor temperature difference ($\Delta T_{\text{o}}$) and the deviation between setpoint and indoor temperature ($\Delta T_{\text{set}}$). Furthermore, interaction terms between the outdoor conditions and control variables were added to explicitly consider the physical coupling effect of the system. Specifically, the product of the absolute humidity and airflow rate was input to reflect the mass conservation law, in which the moisture inflow is proportional to the air mass flow rate, and the interaction between the setpoint temperature and outdoor humidity was incorporated to structure the nonlinearity of the latent heat loads. Consequently, a system was established that could predict the next-day indoor relative humidity profile in advance based on these multidimensional input variables according to the outdoor and operating conditions.

The prediction model employed the XGBoost algorithm, which is a gradient boosting algorithm. To prevent overfitting and induce stable learning convergence, the key hyperparameters were optimized with 300 estimators, a maximum depth of 5, and a learning rate of 0.05. However, owing to the characteristics of time-series data, the recursive forecasting approach, which uses the prediction value from the previous time step as the input for the next step, carries the risk of error accumulation (error propagation) over time, degrading the reliability of long-term simulations. Because the moisture content of indoor air has inertia that does not change abruptly, auto-regressive terms are advantageous for short-term prediction; however, they contribute to error amplification in the long term. To address this error-propagation problem, this study applied a weighted ensemble strategy. This approach constructs Model A, which is trained only with exogenous variables without autoregressive terms, as shown in Fig. 3, and Model B, which includes autoregressive variables to enhance the short-term tracking performance, and then combines the predictions of both models at an optimal ratio. Through this approach, a balance was achieved between the stability of the predictions based on exogenous variables and the dynamic sensitivity of the autoregressive model, thereby inducing stable learning without error divergence, even in long-term simulations.

Furthermore, during the simulation process, the indoor and outdoor humidities were assumed to have reached equilibrium during the nighttime when the AHU was not operating, and the indoor humidity at the start of each day's operation was initialized to outdoor conditions, thereby minimizing errors in the daily enthalpy calculations. This reset mechanism prevented cumulative errors from propagating across multiple simulation days and ensured robust long-term prediction performance.

### 3.3 Cooling load modeling

The core control strategy of this study was to calculate the maximum available capacity according to chilled water supply temperature changes within the rated performance range of the system, and to determine the optimal chilled water temperature by comparing it with the required load. Therefore, it is essential to construct a model that can simulate the dynamic load patterns and magnitudes in buildings. Particularly, in systems utilizing TES, the chilled water supply temperature increases as the stored thermal energy is depleted toward the end of operation, making the energy behavior analysis in that period and the daily cumulative load prediction accuracy key factors determining the success or failure of the system control. This study adopted an RC network load model that offers physical interpretability and computational efficiency [33,34]. Specifically, by following the model structure proposed in prior research [35], both the indoor temperature and the required heat quantity can be simultaneously estimated within a single model with respect to the control input of the setpoint temperature.

The load model was configured based on an RC network, as shown in Fig. 4. The received setpoint temperature and operation signals were used as control variables, and the outdoor temperature, solar radiation, and internal heat gains were used as disturbance variables, reflecting the actual control environment. Through these input variables, the model outputs the indoor temperature and the required heat quantity that the system must supply to maintain the setpoint temperature. This model is based on a 3R2C structure consisting of two capacitances and three physical resistances and is extended to a 4R2C structure by adding a virtual variable thermal resistance $R_{AHU}$ connecting the setpoint and indoor temperatures. The optimization solver derives parameters by configuring an objective function that aims to determine the physical parameters ($R$, $C$) and regression coefficients ($a, b_1, b_2, c$) constituting $R_{AHU}$ that minimizes the indoor temperature and heat quantity prediction errors.

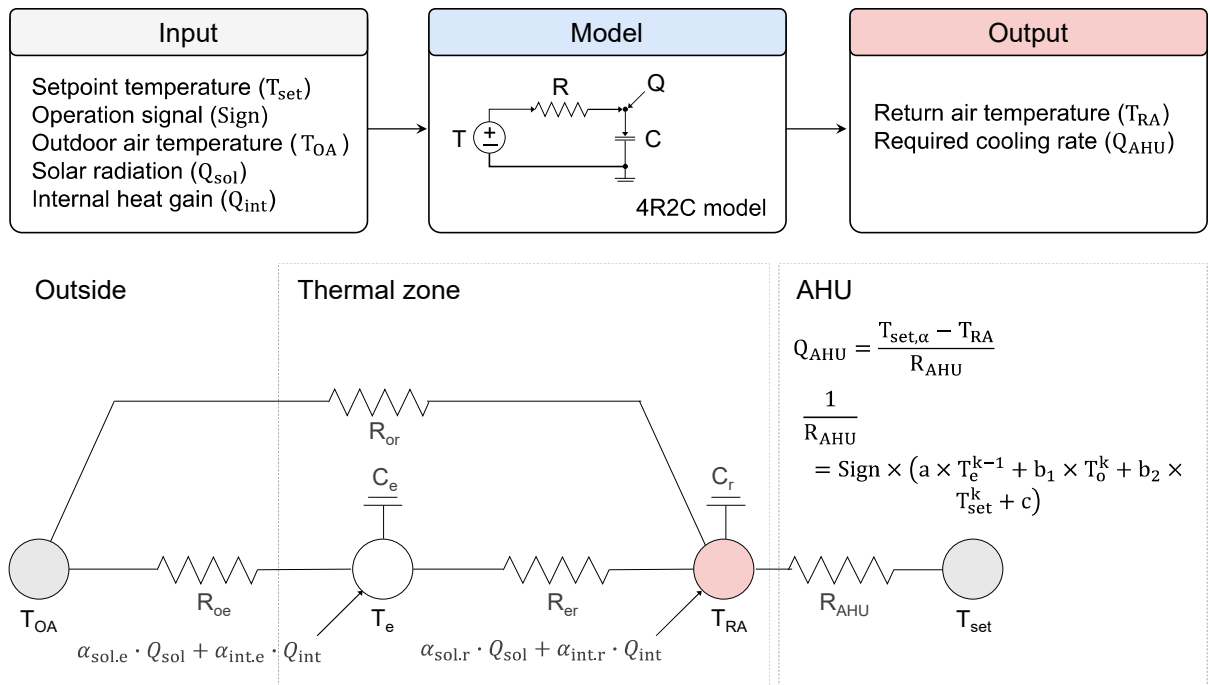

Fig. 4. RC model structure and input and output parameters

### 3.4 Heat exchanger modeling

The cooling performance of a heat exchanger is not a fixed value but varies dynamically according to the inlet conditions of chilled water and air, thus requiring measurements of both the air-side and water-side inlet and outlet conditions for accurate performance analysis. However, the target building in this study measured the air-side return air temperature, relative humidity, and supply air temperature, whereas the heat source side (chilled water side) lacked flow meters and outlet temperature sensors, making only the inlet temperature and valve opening position information available.

In addition, because of the nonlinear flow characteristics of the valves, it is difficult to assume that the valve-opening position is directly proportional to the flow rate. Particularly under low-load conditions, when the valve opening is low, the inlet-outlet temperature difference is minimal, causing problems in which the sensor resolution limits or measurement errors are amplified.

Therefore, to overcome these measurement and observability limitations, this study proposes a physics-informed machine learning (PIML) approach, as shown in Fig. 5, to overcome the limitations of single-side measurement environments and estimate the coil performance using only available data. Such physics-data hybrid approaches can achieve high generalization performance with less data than pure data-driven models [36,37]. Specifically the operating range where valve opening is 95% or above is assumed to be rated operation, and measured effectiveness ($\varepsilon_{m\,ea}$) is calculated. Through this rated-operation filtering process, a reference overall heat transfer coefficient ($UA_{ref}$) is derived, and basic effectiveness ($\varepsilon_{physis}$) following physical structure is calculated using the ε-NTU relationship. Finally, a machine-learning model learns the difference between the theoretical and actual values occurring under various operating conditions, such as the return air temperature, chilled water temperature, and relative humidity, to correct the effectiveness.

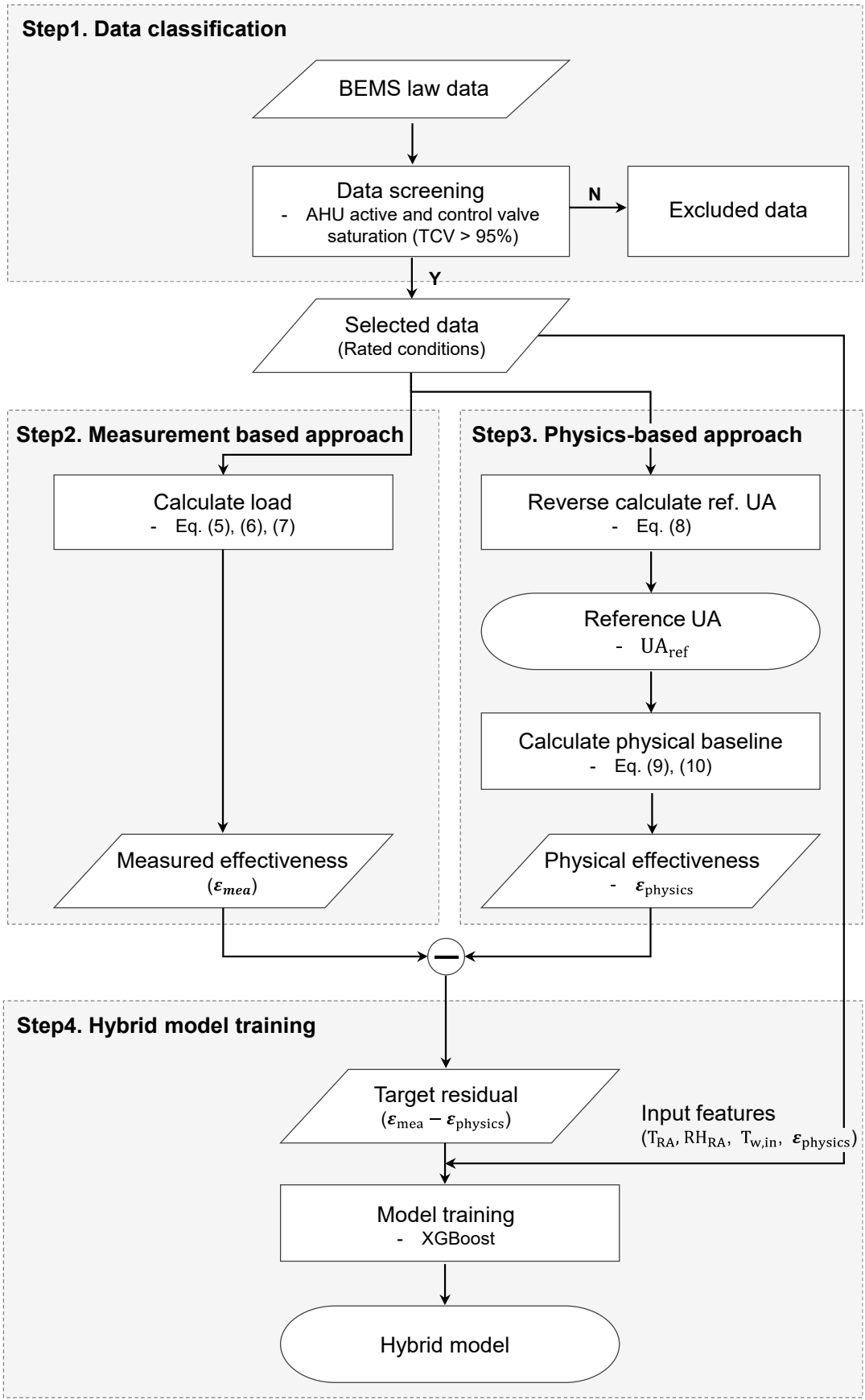

Fig. 5. The process flowchart of the hybrid effectiveness prediction model

In Step 1, the data where the AHU was operating and the valve opening was 95% or above were filtered. Subsequently, the thermodynamic relationships among the chilled water temperature, supply air temperature, and return air temperature were examined to select only the data satisfying physical consistency.

In Step 2, the measured effectiveness was calculated using the selected data. The coil heat-transfer rate must be calculated based on the air-side enthalpy changes because the flow rate and outlet temperature on the water side are unknown. Because the target system was a constant air volume (CAV) type, it was assumed to operate at the design airflow rate when the supply fan was running. Additionally, because humidity sensors exist only on the return air (RA) side, making it difficult to directly calculate the latent heat load on the supply side, the sensible heat ratio (SHR) was fixed at 0.8, consistent with reported values for typical cooling-coil dehumidification behavior in similar climatic regions. The total heat load ($Q_{est}$) was estimated from the sensible heat load, as shown in Equation (5). Maximum exchangeable heat quantity ($Q_{max}$) was calculated using enthalpy difference as shown in Equation (6), and through this calculation, measured effectiveness ($\varepsilon_{mea}$) was derived as shown in Equation (7).

where $Q_{est}$ is the estimated total heat load, SHR is the sensible heat ratio, and $Q_{max}$ is the maximum thermodynamic heat exchangeable by the heat exchanger. $\varepsilon_{mea}$ represents the effectiveness based on measured data; subscripts "in, out" refer to inlet and outlet, respectively, while "air" refers to the air side.

$$Q_{est} = \frac{\dot{m}_{air} \cdot c_{air} \cdot (T_{RA} - T_{SA})}{SHR} \tag{5}$$

$$Q_{max} = \dot{m}_{air} \cdot (h_{a,in} - h_{s,w}) \tag{6}$$

$$\varepsilon_{m\,ea} = \frac{Q_{est}}{Q_{max}} \tag{7}$$

In Step 3, the measured effectiveness ($\varepsilon_{m\,ea}$) calculated under rated conditions and the reference overall heat transfer coefficient ($UA_{ref}$) under rated operation are back-calculated according to Equation (8), and used to calculate predicted effectiveness ($\varepsilon_{physis}$) following physical relationships according to the ε-NTU relationship in Equations (9) and (10).

Here, $UA_{ref}$ is the reference overall heat transfer coefficient under rated flow conditions, and $\varepsilon_{physis}$ is the effectiveness calculated by the physics model.

$$UA_{ref} = -\hbar\,(1-\varepsilon_{m\,ea})\cdot C_{air}\cdot\frac{\dot{m}}{c_{p,m}} \tag{8}$$

$$\varepsilon_{physis} = 1 - \exp(-NTU) \tag{9}$$

$$NTU = \frac{UA_{ref}}{C_{air}} \times \frac{\dot{m}}{c_{p,m}} \tag{10}$$

Finally, in Step 4, the effectiveness predicted from Step 3 is corrected using a machine-learning model. Since coil performance varies according to inlet air and water conditions, the effectiveness ($\varepsilon_{physis}$) calculated with a fixed heat transfer coefficient serves as a baseline in situations where flow rate is not measured. XGBoost, a decision-tree-based ensemble algorithm, was applied as a residual correction model to adjust the performance deviations under actual operating conditions ($T_{RA}, RH, T_{w,h}$). This model learns the difference between model-predicted effectiveness ($\varepsilon_{physis}$) and actually measured effectiveness ($\varepsilon_{m\,ea}$) as the target variable. Input features totaled four, including return air temperature ($T_{RA}$), relative humidity (RH), and chilled water inlet temperature ($T_{w,in}$), which directly affect coil heat transfer characteristics, along with effectiveness ($\varepsilon_{physics}$) derived from the physics model.

In particular, because the filtering process to select only data satisfying the rated operating conditions limits the amount of data available for training, hyperparameters were conservatively set to prevent overfitting and ensure model generalization performance. The maximum tree depth was limited to three to reduce the model complexity, and the learning rate was set to 0.1, with 100 decision trees. The objective function applies a standard regression approach to minimize the mean squared error. This machine learning model has the input–output structure shown in Fig. 6, and learns nonlinear errors that are difficult to explain using physical equations alone, such as partial load characteristics, scale formation, and performance degradation due to aging.

The objective function is defined in Equation (11), which optimizes the model parameters to minimize the difference between the actual and model-predicted values. Through these nonlinear error patterns, which are difficult to explain with physical equations alone, such as partial load characteristics, scale formation, and performance degradation due to aging, the final effective prediction performance is improved.

$$J = \frac{1}{n}\sum_{i=1}^{n}((\varepsilon_{mea,i} - \varepsilon_{physics,i}) - \hat{y}_i)^2 \qquad (11)$$

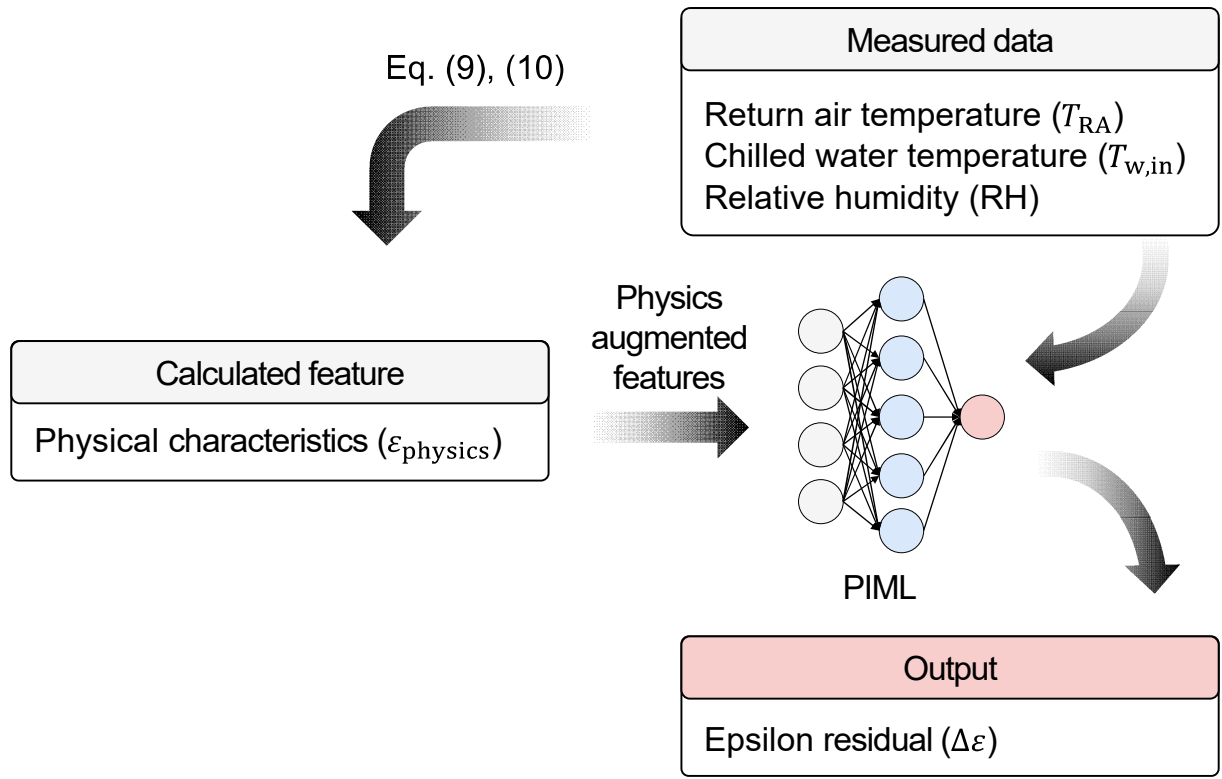

Fig. 6. PIML model input and output parameters

Notably, the proposed model has a methodological distinctiveness in that the explanatory variables for performance prediction are composed of predictable thermodynamic state variables, such as temperature and humidity, rather than real-time operating data, such as the flow rate or valve opening position. The key input variables of the model, the return air temperature ($T_{RA}$) and indoor relative humidity (RH), can be derived in advance using the load and humidity models described earlier based on weather forecast data. Because of this structural configuration, the model can proactively simulate the physical heat exchange limit of the coil even before the actual system operation (pre-operation phase), providing a critical foundation for the integrated framework to perform day-ahead optimization to determine the optimal thermal storage temperature for the next day.

## 3.5 Thermal energy storage modeling

The outlet temperature available during the TES discharge process was determined by the lower-layer temperature according to the internal stratification. Physically rigorous simulation requires the finite volume method (FVM) with numerous nodes or computational fluid dynamics (CFD) analysis; however, these impose high computational loads, limiting their use as optimization or control decision-making tools. The purpose of this study is not to analyze the microscopic and complex fluid behavior inside the TES but to determine the optimal chilled water supply temperature suitable for load handling considering the physical limits of the system and heat pump efficiency characteristics. Therefore, the modeling focus of this study was on verifying the load-response capability rather than the time-series temperature tracking. Specifically, predicting the outlet temperature at the end of operation according to the cumulative discharge heat and ensuring chilled water supply stability during peak periods when thermal stress is the highest were defined as the core model requirements. Therefore, this study modeled the TES as a dynamic system with a single-state variable using discrete-time state-space equations based on the energy conservation law. The state variable $x$ represents the internal energy (average temperature) of the TES, the input $u$ is the discharge load, and the output $y$ is calculated using the correlation equations derived from the measured data. The state equation representing the thermal behavior of the TES is given by Equation (12). This equation simulates the dynamic characteristics, where load accumulation over time leads to an increase in the average temperature inside the TES.

where $T_{\text{avg}}(t)$ is the remaining energy inside the TES at time $t$, $Q_{\text{bad}}$ is the load, M is the mass of water in the TES, and $C_{p,\text{water}}$ is the specific heat of water.

$$T_{\text{avg}}(t+1) = T_{\text{avg}}(t) - \frac{Q_{\text{bad}}(t)}{M \cdot C_{p,\text{water}}} \cdot \Delta t \tag{12}$$

While the average temperature of the TES is calculated at each time step using the state equation in Equation (12), the actual discharge temperature supplied to the coil exhibits a different behavior from the average temperature owing to stratification effects. Generally, low temperatures are maintained during the early discharge phase; however, the outlet temperature increases nonlinearly as the storage capacity is depleted. In this study, the characteristic curves defining the relationship between the average TES temperature and the discharge temperature were used to simulate the stratification behavior of the TES.

The TES outlet temperature has very different flow characteristics when the stratification is maintained and when it collapses at the critical point, making predictions using a single limited model. Therefore, this study divided the TES behavior into two cases corresponding to the most important prediction points: the peak phase and the end of the operation phase, and established models optimized for each.

First, the outlet temperature at the peak load point was simplified using actual operational data to express the correlation between the two variables in a third-order polynomial form, as shown in Equation (13), which was utilized as the output equation. This method can effectively capture the point at which cooling limits are reached by simulating the supply temperature increase according to the load accumulation without complex fluid analysis. Thus, this equation functions in the present simulation as a tool for verifying the validity of the proposed framework by simulating the general discharge pattern of a stratified TES, rather than for the precise prediction of a specific TES configuration.

$$T_{\text{out}}(t) = a \cdot T_{\text{avg}}^{3} + b \cdot T_{\text{avg}}^{2} + c \cdot T_{\text{avg}} + d \tag{13}$$

The final discharge temperature at the end of the operation, which ultimately determines the operational feasibility of the TES, was determined by considering the cumulative discharge quantity and operating time, as shown in Equation (14). Unlike existing models that use only simple average temperatures, this model more accurately captures the outlet temperature rise characteristics at the end of the discharge by considering the interaction effect between the discharge rate and cumulative load.

Here, $T_{avg}$ is the current average temperature, $\Delta T$ is the temperature rise from the initial temperature, and $LR$ is the load rate defined as cumulative discharge divided by operating time.

$$T_{out,end} = a \cdot T_{avg}^2 + b \cdot T_{avg} + \Delta T + d \cdot LR + e \cdot (\Delta T \cdot LR) + f \tag{14}$$

## 4. Target Case

### 4.1 Zone and AHU

The target building was an office building located in Seoul that was completed in 2007. The building was used as an office facility, and one office floor was selected as the reference floor for analysis. Although the actual TES of the target building supplies cooling to multiple air-conditioned zones in an integrated manner, this study simulated the entire system based on a reference floor model to explain the operating process of the proposed decision-making framework. Accordingly, the total building cooling load, which was the input condition for the TES model, was calculated by scaling based on the reference floor load pattern. Specifically, the ratio between the total rated cooling coil capacity of the AHUs installed throughout the building and the reference floor AHU coil capacity was calculated and used as a weighting factor to derive the total load. That is, assuming that the load patterns of the other zones in the building were similar to the thermal behavior of the reference floor, a simple summation approach proportional to the total equipment capacity ratio was adopted.

The analysis scope was set to the selected AHU and indoor space, and modeling was performed by focusing on the operating characteristics of the HVAC system and the indoor environmental conditions within the scope. Each floor is equipped with an AHU operating supply and exhaust fans in constant air volume mode, with the specifications shown in Table 1.

Table 1. Building and system properties

| AHU information | | | Value | Unit |
|---|---|---|---|---|
| AHU | Cooling capacity | | 194 | kW |
| | Fluid flow rate | | 33,600 | kg/h |
| | Air flow rate | SA | 32,820 | CMH |
| | | RA | 30,900 | CMH |
| | Fluid temperature | Inlet | 7 | °C |
| | | Outlet | 12 | °C |

The target building had a building automation system (BAS) installed to monitor and store various temperature data, as listed in Table 2, with the measurement locations indicated in Fig. 7. All data were measured at 1-minute intervals, this study preprocessed the temperature and signal data at 5-minute intervals. The average value was used after excluding the upper and lower 10% of the five temperature data points collected per minute as outliers, and operation signals were considered active if three or more of the five minutes indicated an operation.

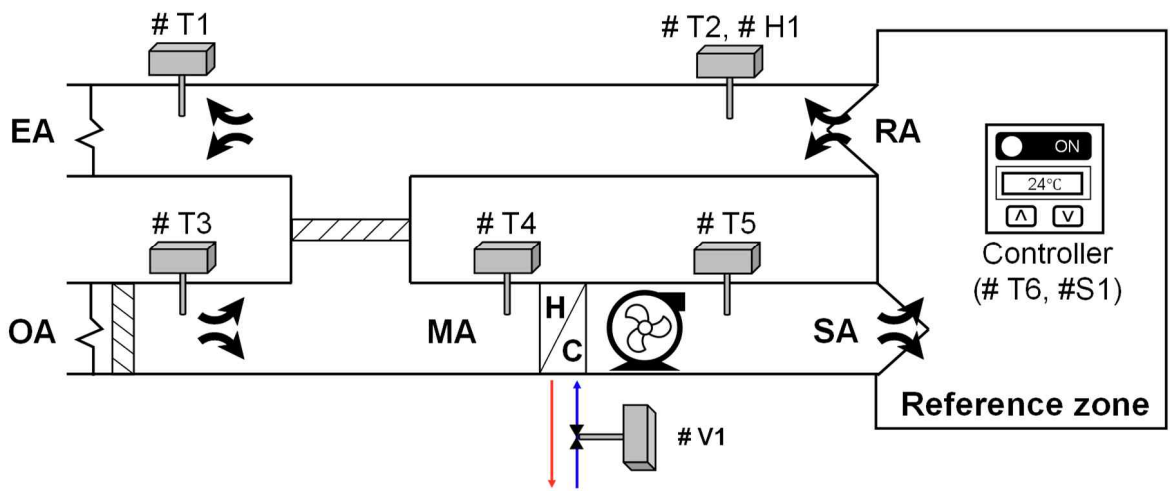

Fig. 7. Schematic diagram of AHU configurations of the target zone

Table 2. BAS Measurement point

| Parameter | Description | Unit | Sampling period | Collection type |
|---|---|---|---|---|
| #T1 | Exhausted air temperature | °C | 5min | Timestamp |
| #T2 | Return air temperature | °C | | |
| #T3 | Outdoor air temperature | °C | | |
| #T4 | Mixing air temperature | °C | | |
| #T5 | Supply air temperature | °C | | |
| #T6 | Setpoint temperature | °C | | |
| #H1 | Return air humidity | % | | |
| #S1 | Supply fan sign | - | | |
| #V1 | Valve position | % | | |
| #T7 | Chilled water supply temperature | °C | | |
| #T8 | Chilled water return temperature | °C | | |
| #TE3 | Upper layer temperature | °C | | |
| #TE15 | Lower layer temperature | °C | | |

## 4.2 Thermal storage and hydronic system

The target building employs a water thermal storage system that produces and stores cooling energy using off-peak nighttime electricity. The overall configuration of the heat source and hydronic system is shown in Fig. 8. The system primarily consists of a chiller, TES tank, heat exchanger, chilled water, and cooling water circulation pumps. The main design specifications of the TES are listed in Table 3. Diffusers were installed at the top and bottom to effectively maintain the internal stratification.

Table 3. Design specifications of TES

| Item | Value | Unit |
| --- | --- | --- |
| Charging material | Water | - |
| Total charging capacity | 2,286 | kW |
| Tank net volume | 273 | $m^3$ |
| Max temperature | 13 | °C |
| Min temperature | 5 | °C |

During the night, chilled water produced by the chiller is directly supplied to the bottom of the TES, while relatively warmer water from the top of the TES flows into the chiller's evaporator to be cooled, thereby performing the charging operation. During daytime cooling, low-temperature chilled water from the bottom of the TES is supplied to the primary side (TES side) of the heat exchanger owing to the temperature stratification inside the tank. This low-temperature water exchanges heat with the secondary side (load side) and circulates through the heat exchanger to supply cooling to the load side. The chilled water on the secondary side was supplied to the AHU coil to handle the indoor cooling loads, and then returned to the heat exchanger at an elevated temperature. The water that absorbed heat from the secondary side while passing through the primary side of the heat exchanger returned to the top of the TES, maintaining a stable temperature stratification inside the tank. Flow meters and temperature sensors were installed at key system points, as shown in Fig. 8 and Table 2, and were utilized for real-time operation monitoring and control. In this study, it was assumed during the simulation that the low-temperature chilled water delivered from the primary side was equally transferred to the secondary side via heat exchange.

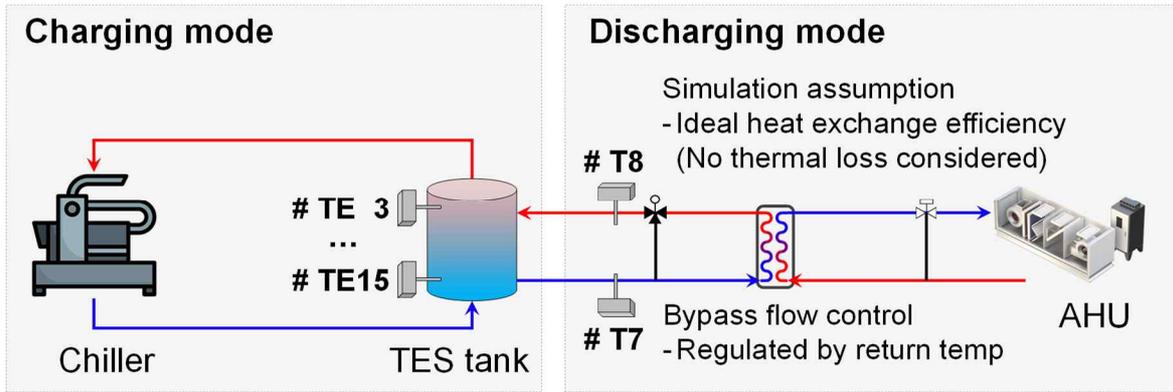

Fig. 8. Schematic diagram of AHU configurations of the target zone

## 5. Results

**5.1 Performance validation of indoor relative humidity prediction model**

The chilled water supply limit temperature determination framework proposed in this study requires an accurate simulation of the thermodynamic behavior of the coil. The enthalpy information of the coil inlet air, calculated from the indoor temperature and relative humidity, is essential for this purpose. Therefore, the performance of the humidity prediction model developed using summer data (July–September) was evaluated. A performance evaluation was conducted for the two-week HVAC system operating period, separated from the total data, as the test set. As shown in Table 4, the model demonstrated excellent prediction performance, with an average CVRMSE of 7.22% and $R^2$ of 0.78. Figs. 9 and 10 present the measured indoor humidity and model prediction results as time series graphs and scatter plots, respectively.

The analysis results in Fig. 10 show a tendency for a slightly increased prediction error in low-humidity ranges below 50% relative humidity compared with high-humidity ranges. This is because the training data samples in the low-humidity range were relatively insufficient, as the data in this study were concentrated in the hot and humid summer seasons. From the perspective of cooling energy optimization, which is the core purpose of this study, the prediction error in this range was judged to have a limited impact on the overall system control performance. This is because in low relative humidity ranges, air enthalpy changes are dominated mainly by sensible heat; therefore, the sensitivity of humidity prediction deviations to the overall enthalpy calculation and cooling load estimation is significantly lower than that in high humidity ranges. Consequently, because the proposed model secures high accuracy in high-humidity ranges where the majority of the cooling load occurs and dehumidification is important, it is appropriate for application to the overall system optimal control.

Table 4. Model comparison for RH

|       | NMBE (%) | CVRMSE (%) | $R^2$ |
|-------|----------|------------|-------|
| Week1 | 2.89     | 8.12       | 0.82  |
| Week2 | 2.94     | 6.32       | 0.74  |
| Ave   | 2.92     | 7.22       | 0.78  |

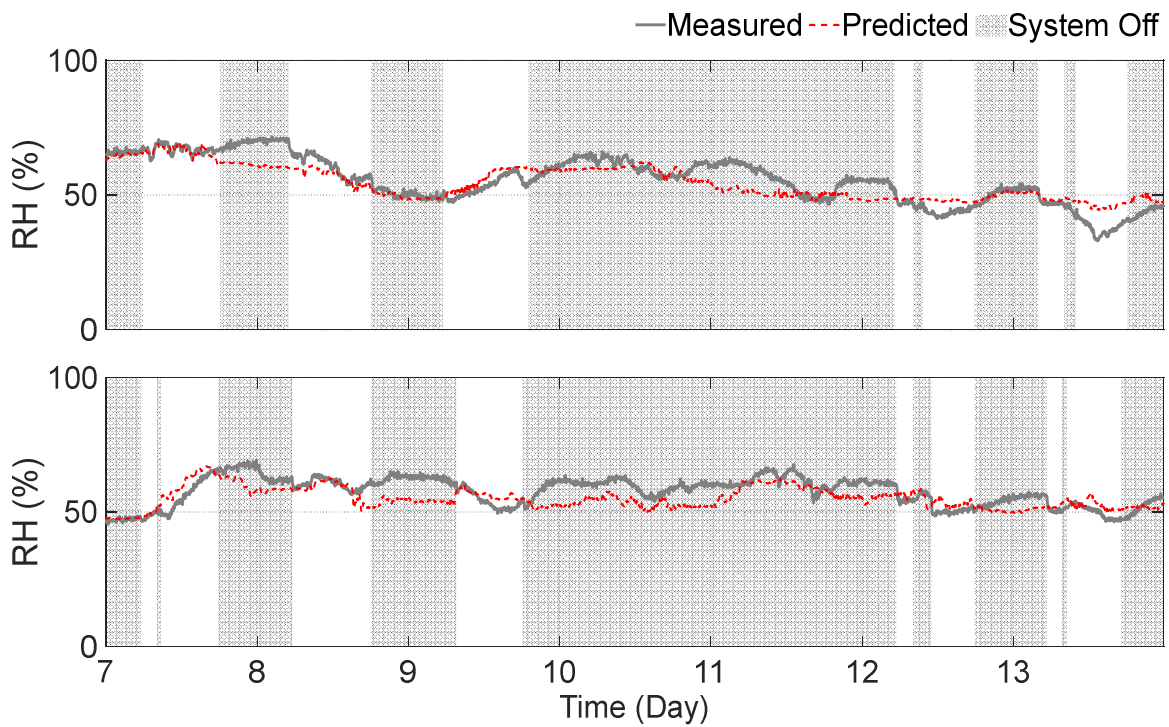

Fig. 9. Comparison of measured and predicted indoor relative humidity

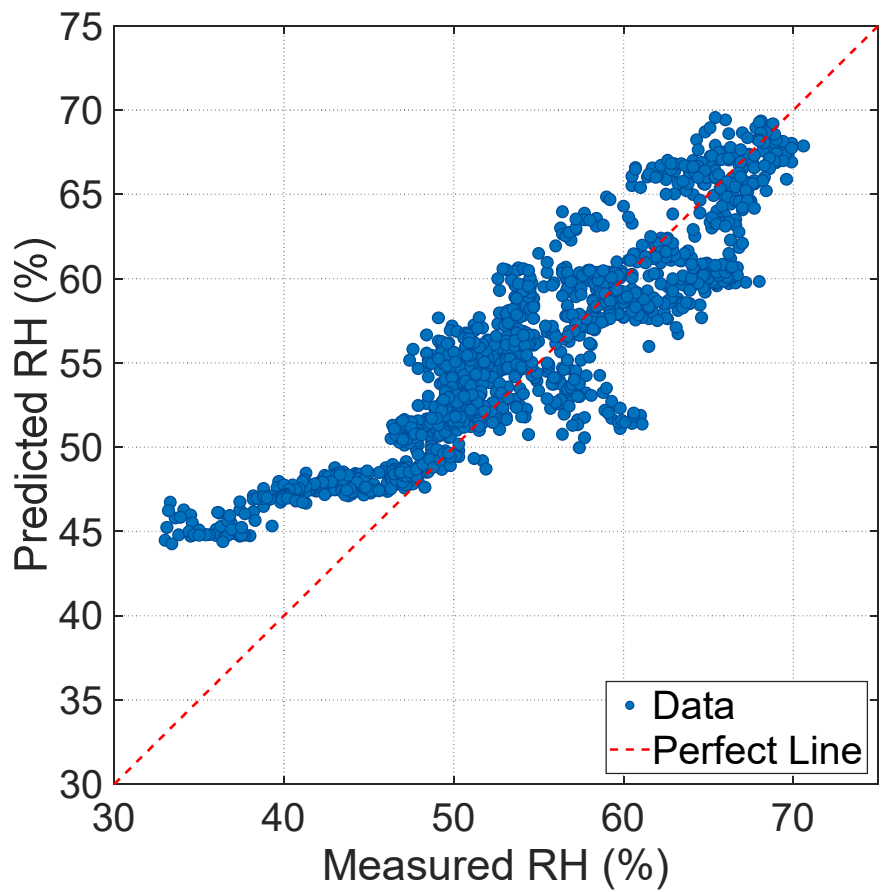

Fig. 10. Correlation between measured and estimated relative humidity values

The developed prediction model precisely tracked humidity change patterns due to daytime AHU operation and indoor humidity patterns due to outdoor humidity variations. This high prediction performance is attributed to the effective reflection of humidity time-series continuity through the introduction of autoregressive input variables (lagged variables) that reflect past state information in current predictions. In addition, the nonlinear learning capability of the XGBoost algorithm was determined to appropriately consider the influence of complex environmental variables. The validated humidity prediction model was utilized as the input data for the subsequent heat exchanger model, forming the basis for ensuring the reliability of the enthalpy-based coil load calculation and the optimal chilled water temperature derivation.

**5.2 Performance validation of cooling load model**

The cooling load required in each zone to maintain the set target temperature must first be calculated to determine the optimal chilled water supply temperature, which is the goal of this study. The cooling load is calculated using heat balance equations based on an RC model that considers the difference between the current indoor temperature and the setpoint temperature and the thermal inertia of the building. Therefore, for the RC model to produce reliable results, the temperature change pattern that appears when cooling energy is supplied to an indoor space must be similar to the actual measured data. This indicates how well the model simulates the actual thermal inertia of the building. This study verified the prediction accuracy of the RC model using the tracking performance of the measured indoor temperature and cooling load as evaluation metrics.

The RC model was calibrated using the initial two weeks of data for key parameters, and the performance was evaluated using the subsequent four weeks of data as the test set. The indoor temperature prediction performance evaluation results showed that the model accurately predicted the indoor temperature according to the set-point temperature, as shown in Fig. 11. As shown in Table 5, the CVRMSE was calculated as 2.36%, confirming the precise tracking of the measured indoor temperature. In contrast, $R^2$ was very low at 0.1. This is because, as shown in Fig. 11, the actual measured data show severe oscillation owing to the continuous controller operation to maintain the setpoint temperature, whereas the developed load model does not include such detailed control operations and shows stable behavior near the setpoint temperature without oscillation. This difference in short-term variability is thought to be the cause of the low $R^2$.

Table 5. Model comparison for return air temperature

|  | NMBE (%) | CVRMSE (%) | $R^2$ |
|---|---|---|---|
| Week1 | -0.83 | 2.53 | -0.07 |
| Week2 | 0.3 | 2.03 | 0.35 |
| Week3 | 0.14 | 2.22 | 0.41 |
| Week4 | -0.57 | 2.65 | -0.3 |
| Ave | -0.24 | 2.36 | 0.1 |

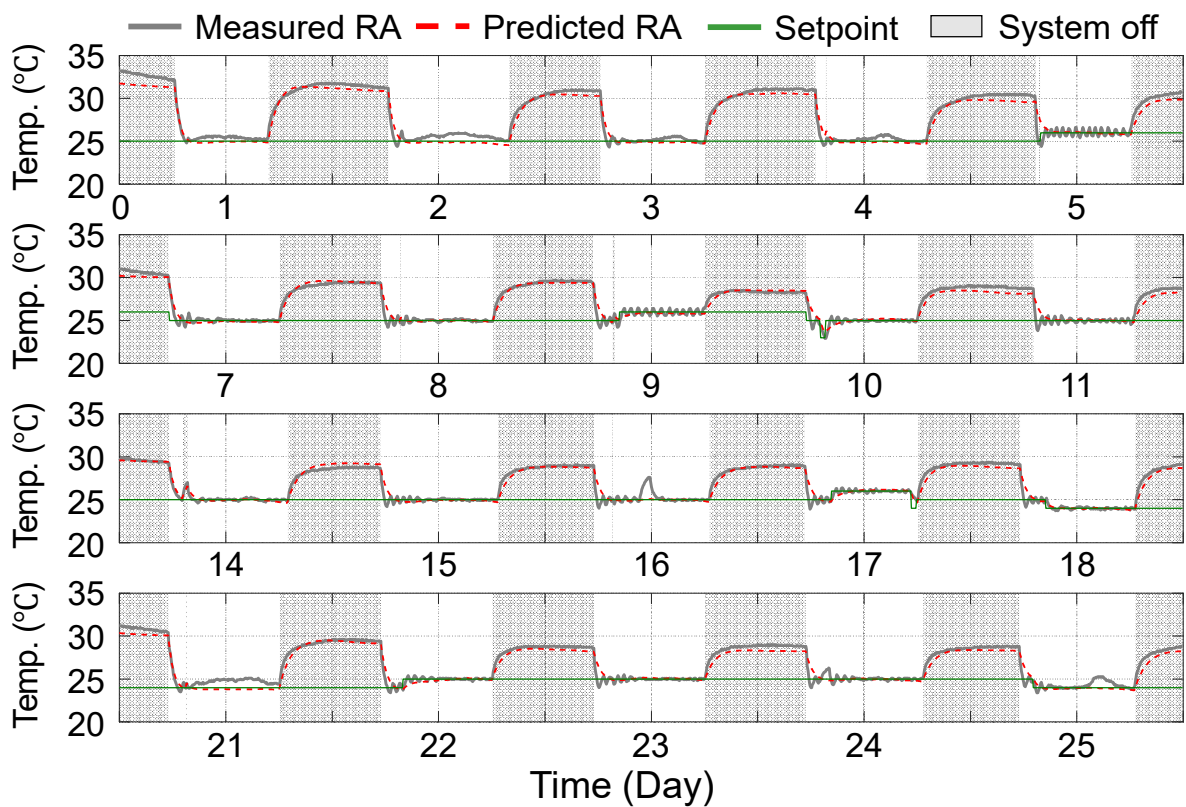

Fig. 11. Comparison of measured and predicted return air temperature

Next, the cooling load prediction results were compared with the measured values. The main purpose of analyzing the cooling load in this study is not to precisely match instantaneous control variations but to determine the chilled water supply limit temperature, which serves as the criterion for judging cooling feasibility considering the physical limits of the hydronic system and TES temperature rise. Therefore, when evaluating the cooling load, in addition to timestep-by-timestep prediction values, a cumulative value comparison showing the period-by-period total prediction performance was conducted in parallel.

The time-step-by-time-step performance analysis results showed a CVRMSE of 39.35% and $R^2$ of 0.18, as shown in Table 6, with $R^2$ appearing similarly low as the indoor temperature prediction results. As confirmed in the timestep-by-timestep cooling load comparison graph in Fig. 12, this error occurs because the model fails to track the severe control oscillation of the actual system. However, the cumulative cooling load graph confirmed that the model accurately predicted the total required load. For the cumulative data, the CVRMSE improved significantly to 17.13% compared to the timestep-by-timestep results, and $R^2$ appeared to be very high at 0.92. This indicates that although the model may not perfectly simulate instantaneous control variations, it has a very high correlation with the actual system in terms of long-term energy consumption trends and cumulative totals, making reliable predictions.

Table 6. Model comparison for cooling load

|  | Instantaneous cooling load | | | Cumulative cooling load | | |
| --- | --- | --- | --- | --- | --- | --- |
|  | NMBE (%) | CVRMSE (%) | $R^2$ | NMBE (%) | CVRMSE (%) | $R^2$ |
| Week1 | 7.88 | 31.22 | 0.15 | 6.77 | 14.06 | 0.95 |
| Week2 | -3.17 | 43.17 | 0.29 | -5.6 | 15.96 | 0.93 |
| Week3 | 17.39 | 42.64 | 0.07 | 16.87 | 22.31 | 0.88 |
| Week4 | 10.2 | 30.39 | 0.21 | 12.95 | 16.2 | 0.94 |
| Ave | 8.08 | 39.35 | 0.18 | 7.75 | 17.13 | 0.92 |

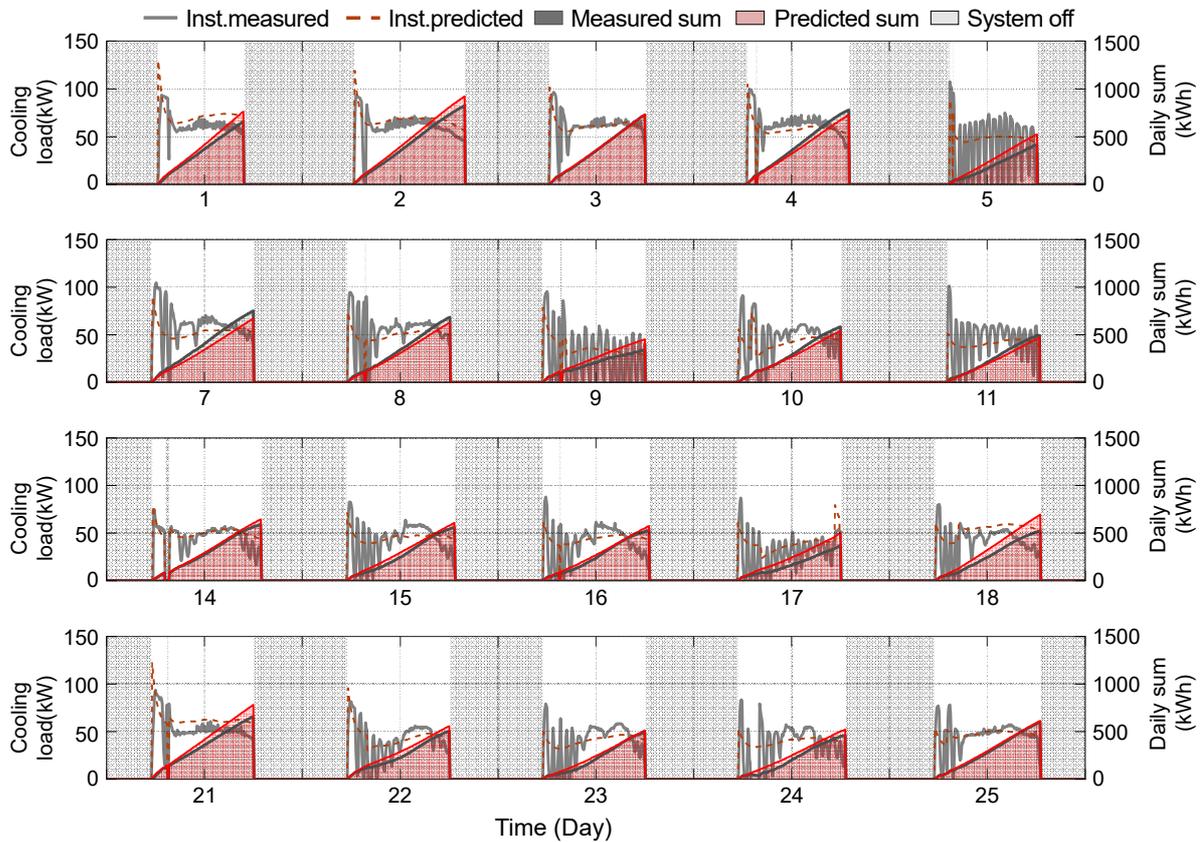

Fig. 12. Comparison of measured and predicted instantaneous and cumulative cooling load

Despite instantaneous control errors, because the overall load trends and period-by-period cumulative energy consumption prediction performance were excellent, the model was judged to have sufficient validity for calculating the chilled water supply limit temperature, which was the main purpose of this study. The cooling load data secured in this manner were connected as target values for the AHU coil model and utilized as data for back-calculating the optimal chilled water supply temperature for handling the load.

**5.3 Performance validation of cooling coil model**

This section derives the theoretical upper limit of the chilled water supply temperature that enables cooling, which is determined solely by the indoor load and coil heat-transfer performance, regardless of the TES supply capacity. This critical value becomes the quantitative criterion for the optimal discharge temperature that should be targeted during the actual TES operation.

The overall prediction accuracy of this model depends significantly on the accuracy of the reference overall heat transfer coefficient ($UA_{ref}$). In particular, considering the current measurement environment, where partial-load validation is difficult, performance verification in the region approaching the rated conditions is the most reasonable means to demonstrate the physical validity of the model. Accordingly, quasi-nominal data, where the AHU was operating and the valve opening was 95% or above, were used to verify the model's baseline performance, with the results shown in Fig. 13.

This study applied the ε-NTU methodology to evaluate the heat exchange performance of the coil using supply air temperature and cooling capacity as indicators. The analysis results showed R² and CVRMSE of the supply air temperature of 0.85 and 2.84%, respectively, as shown in Table 7, whereas the cooling capacity showed R² of 0.82 and CVRMSE of 9.98%. Although direct partial load validation was not possible owing to the absence of sensors, this signified that the reference overall heat transfer coefficient, a key variable for modeling, was accurately derived. Therefore, this model secured a valid baseline on which physical scaling laws can be applied for partial-load operating conditions based on the established reference point.

Table 7. Model comparison for cooling coil

|  | Supply air temperature | | | Cooling coil capacity | | |
|---|---|---|---|---|---|---|
|  | NMBE (%) | CVRMSE (%) | $R^2$ | NMBE (%) | CVRMSE (%) | $R^2$ |
| Hybrid model | 0.15 | 2.84 | 0.85 | -0.51 | 9.98 | 0.83 |

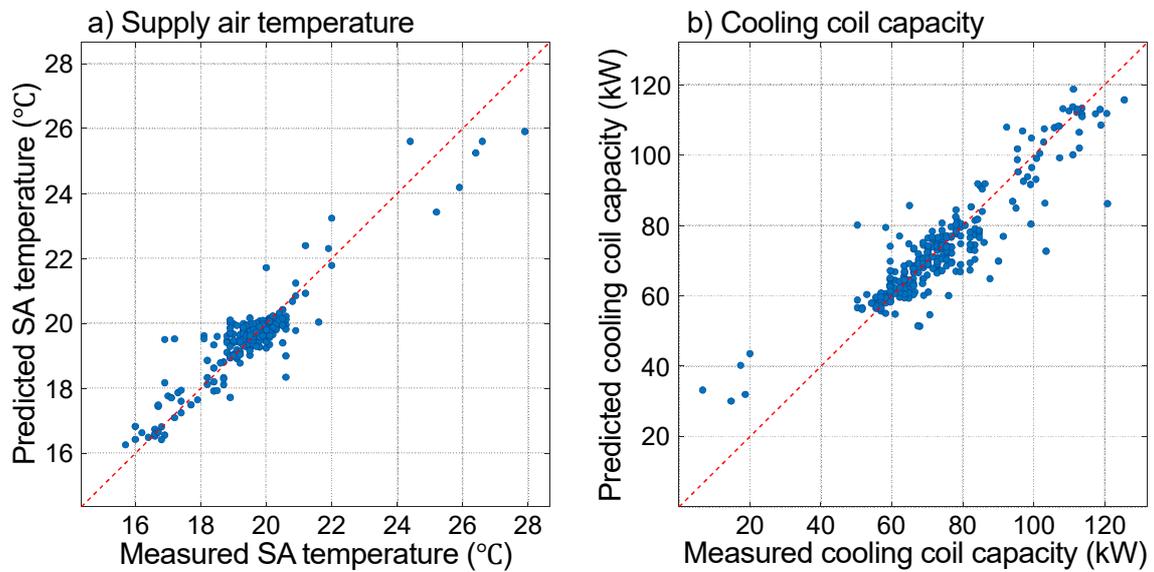

Fig. 13. Comparison of predicted and measured cooling coil performance

The validated coil and load models were linked to simulate the maximum chilled water inlet temperature capable of handling loads. Fig. 14 presents the time-series analysis results for four weeks of operational data. The upper graph shows the control status of the measured indoor and setpoint temperatures, whereas the lower graph compares the measured supplied chilled water temperature with the cooling limit temperature derived in this study. Here, the cooling-limit temperature is not an actual supplied value but represents a guideline as a thermal threshold allowed for controlling the indoor temperature under the load conditions at that time.

The analysis results showed that in some sections of weeks 1 and 4, the actual supply water temperature exceeded the temperature limit suggested by the model, and deviation phenomena in which the indoor temperature increased without tracking the setpoint were confirmed at those points. However, in sections where the actual supply water temperature was maintained below the temperature limit, the indoor temperature was confirmed to be stably controlled within the set-point range.

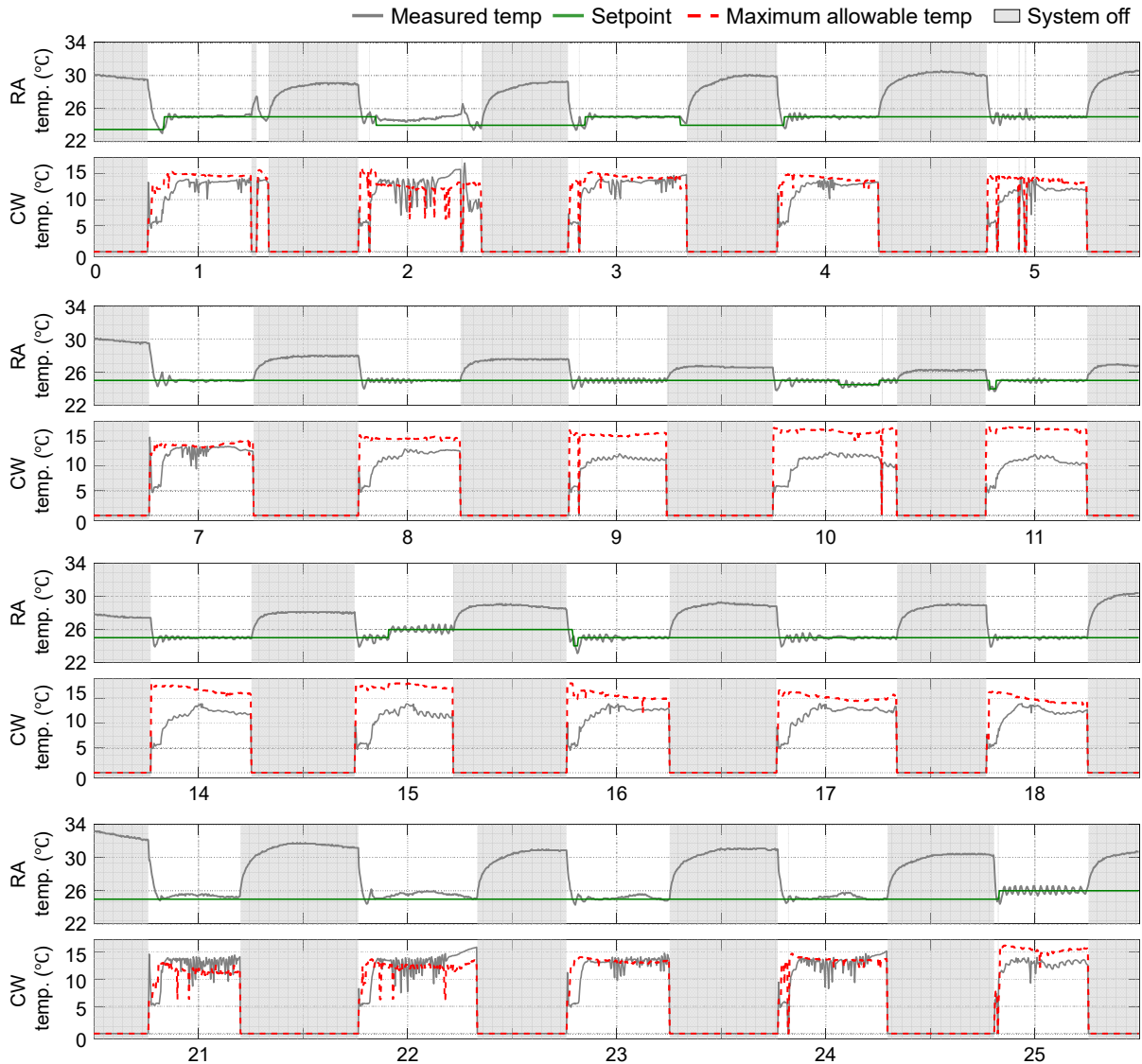

Fig. 14. Simulation results of indoor temperature and chilled water temperature limits

The potential of raising the chilled water supply temperature to improve the heat pump efficiency was evaluated. Here, the potential is defined as the deviation between the chilled water limit temperature and the actual supply temperature. Fig. 15 shows the frequency distribution of potential for the entire analysis period, and analysis results confirmed that an average temperature increase margin of 4.52 °C exists. This result implies that the current operating method supplies chilled water at temperatures lower than necessary and indicates the potential to save energy efficiently by moving the operating point of the heat pump to a high-efficiency range through supply temperature increase control.

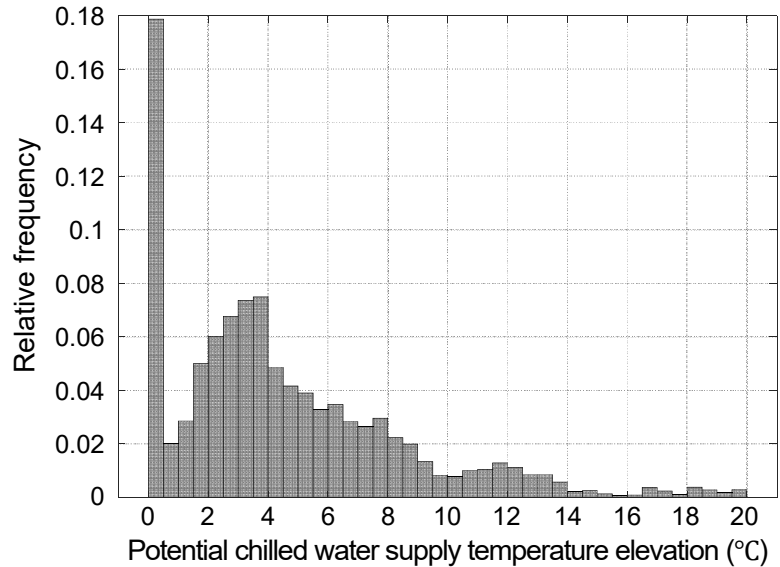

Fig. 15. Histogram of potential for increasing chilled water supply temperature

**5.4 Performance validation of thermal energy storage model**

The core of the proposed integrated control strategy is to determine the operational availability range in advance such that the TES outlet temperature does not exceed the physical critical temperature of the coil. Therefore, the validity verification must precede the precise simulation of the dynamic changes in the TES outlet temperature that accompany the discharge process. Therefore, this study comparatively analyzed the measured and predicted values targeting the peak time when the cooling load is maximum during the operating period and the end of operation when the water temperature reaches its maximum owing to thermal storage depletion, which is critical for ensuring the cooling stability of the system.

Fig. 16 shows a scatterplot of the model prediction performance at these two time points. The analysis results showed a peak time prediction performance with a CVRMSE of 2.78% and R² of 0.91, as shown in Table 8, indicating very precise tracking of the actual TES behavior. This result implies that the model can accurately predict the supply temperature at the point where the loads are concentrated, thereby providing a reliable control criterion. On the other hand, the performance at the end of the operation showed a CVRMSE of 17.84% and R² of 0.79, showing a tendency for increased error compared to the peak time.

Table 8. Validation of discharge temperatures in the thermal energy storage model

| Peak time | | | | | End of operation | | | | |
|---|---|---|---|---|---|---|---|---|---|
| MBE (°C) | NMBE (%) | CVRMSE (%) | $R^2$ | Standard deviation (°C) | MBE (°C) | NMBE (%) | CVRMSE (%) | $R^2$ | Standard deviation (°C) |
| 0 | 0 | 2.78 | 0.91 | 0.3 | 0.001 | -0.01 | 17.84 | 0.79 | 3.11 |

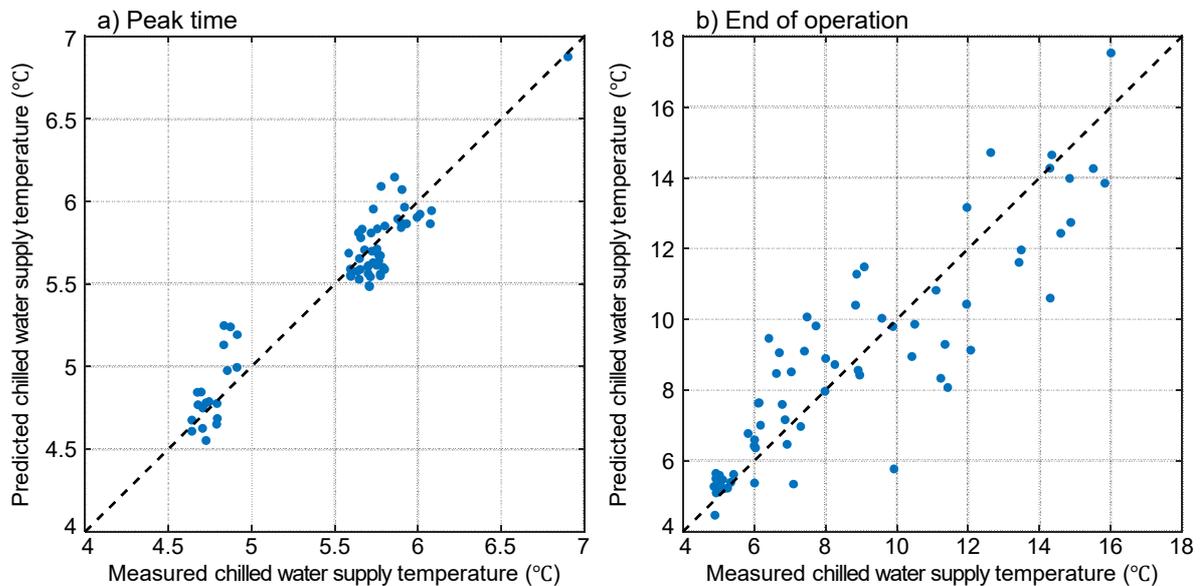

Fig. 16. Comparison of predicted and measured chilled water supply temperature

Fig. 17 presents the Bland-Altman analysis results for validating the TES model, showing the distribution of differences (Y-axis) according to the mean of the two variables (X-axis). As shown in Table 8, the peak load time data clustered near the zero-error baseline with a very low standard deviation (±0.3 °C), suggesting that the model prediction reliability was secured owing to robust stratification inside the TES. On the other hand at the end of operation, thermal uncertainty increased due to stratification collapse and fluid mixing, showing a tendency for data dispersion to expand approximately 10-fold compared to peak (±3.11 °C).

However, despite this increase in dispersion, the mean bias error(MBE) at the end of the operation was very small (0.001). This demonstrates that although the model may not perfectly track the variability caused by the local turbulent behavior, it accurately simulates the macroscopic thermal behavior of the system. Therefore, it was judged to have sufficient validity as a decision-making tool for determining the physical feasibility limits of TES operations.

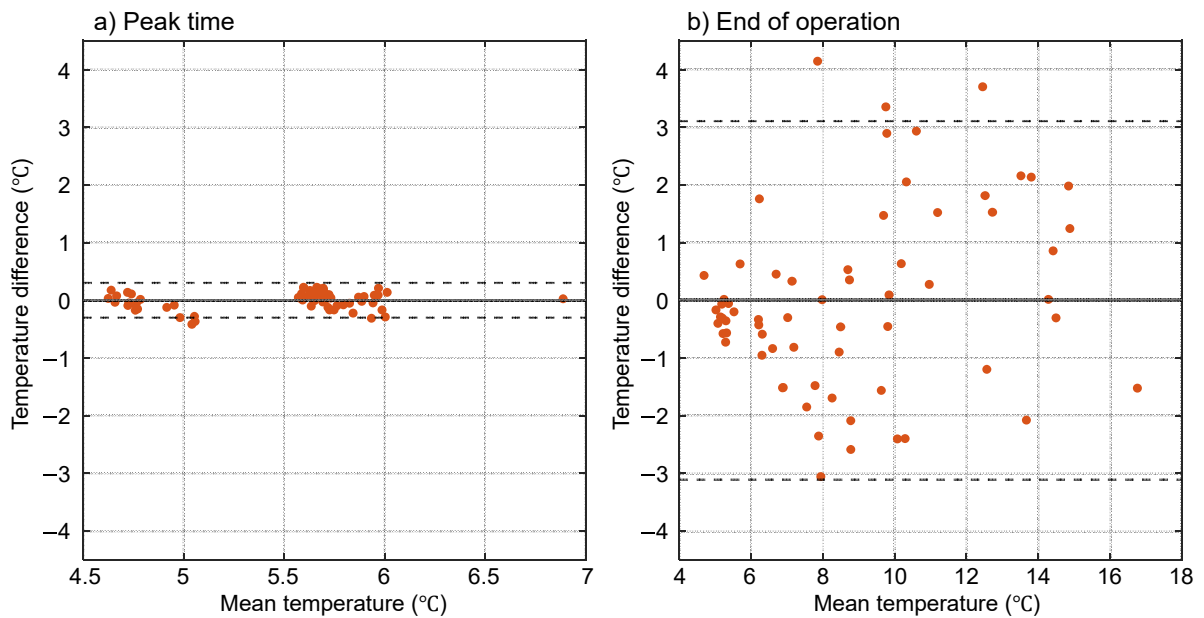

Fig. 17. Bland-Altman plots of measured versus predicted TES discharge temperatures

## 5.5 Simulation results of integrated optimal control strategy

This section integrates previously validated unit models to extend the analysis target to the entire building, and verifies its feasibility at the system level. The simulation was performed under the assumption that the reference floor load pattern was proportional to the entire building (five floors), considering the hourly outlet temperature changes according to the TES discharge characteristics. Thus, the optimal initial charging temperature required to handle the target cooling load was derived, and whether the condition can be implemented within the physical discharge performance limit (physical feasibility) of the TES was evaluated comprehensively. Because the results of this simulation were not actual measured data obtained by operating the building with the corresponding control logic, the performance was analyzed by comparing the simulation results with actual data obtained using conventional methods (fixed setpoint temperature) during the same period. The analysis target period was selected as the summer season (July–August), when cooling energy consumption is the most concentrated throughout the year. This period includes various load variation characteristics from partial to peak loads approaching the system design capacity, making it suitable for comprehensively evaluating the versatility and energy-saving potential of the proposed control strategy. Fig. 18 presents a comparison of the daily operating results between the proposed optimal control strategy and conventional operating method, divided into (a) initial TES temperature, (b) TES temperature at the end of operation, and (c) daily cooling load.

First, the analysis results in Fig. 18(a) confirm that the conventional operating method operates at a constant temperature, regardless of the load magnitude. In contrast, the proposed model showed a pattern of flexibly adjusts the initial TES temperature according to the predicted load. In particular, the difference between the two methods was significant during simulation week two (days 8– 12) when the cooling load was relatively low. The conventional method maintained low initial temperatures by excessively cooling the TES to design reference temperature despite reduced load, whereas the proposed control method increased the initial TES temperature by approximately 2.55 °C compared to conventional operation within the range where there would be no problem handling next-day cooling loads. This means that unnecessary energy was saved, and the potential to increase the operating efficiency of the heat pump was secured.

Considering the thermodynamic characteristic that heat pump COP improves by approximately 2–4% for every 1 °C rise in evaporation temperature [9, 38], the 2.55 °C supply temperature increase secured in this study is theoretically analyzed to lead to efficiency improvement effects of approximately 5–10%. Furthermore the proposed strategy was confirmed to successfully implement a full discharge cycle that utilizes stored cooling without residual by allowing the TES temperature at end of operation to reach the discharge limit of 17 °C.

Fig. 18(b) shows the residual TES temperature at the end of the HVAC operation. During the period from day 8 to day 12 when cooling loads were relatively low, the proposed model showed a pattern where the TES temperature sufficiently rose to approximately 16–17 °C near the discharge limit and finished at end of operation, despite setting the initial TES temperature high at approximately 9 °C. This behavior contrasts with the conventional operating method, which is completed at low temperatures despite low loads, resulting in unused cooling. In other words, the proposed control strategy suggests that the operational efficiency is maximized by minimizing excessive safety margins and fully utilizing the stored cooling within the range of maintaining the chilled water supply capability required for coil inlet enthalpy control, even when considering model prediction errors.

Particularly on day 2, the initial temperatures were similar between the measured and model-predicted values, but a difference of approximately 4 °C occurred at the end of the operation. Day 2 had a high daily cumulative cooling load, a point where the effective thermal capacity of the TES reached its limit and stratification collapsed. A discussion related to the stratification is provided in the appendix.

Therefore, the error on days when the stratification was destroyed was more appropriately interpreted as a discrepancy between the physical stratification collapse phenomenon occurring under load conditions exceeding the design capacity and the idealized stratification model rather than a model performance defect. The main purpose of this study was to determine the margin for raising the TES supply temperature. The analysis of conditions requiring lower temperatures owing to insufficient TES design capacity is beyond the scope of this study.

The simulation results confirmed that the physical tendency of the optimal TES charging temperature could be variably adjusted according to the predicted cooling load magnitude. From a practical perspective, this means that it is possible to establish control strategies that proactively optimize the charging setpoint temperature based on next-day weather and predicted loads. However, empirical methodologies at sites that find optimal values through trial and error have low feasibility considering the complexity of large-scale buildings and the risk of reducing occupant work productivity. Additionally, the value of this framework is that it does not provide analytical data for identifying the causes of performance degradation or systematic improvement.

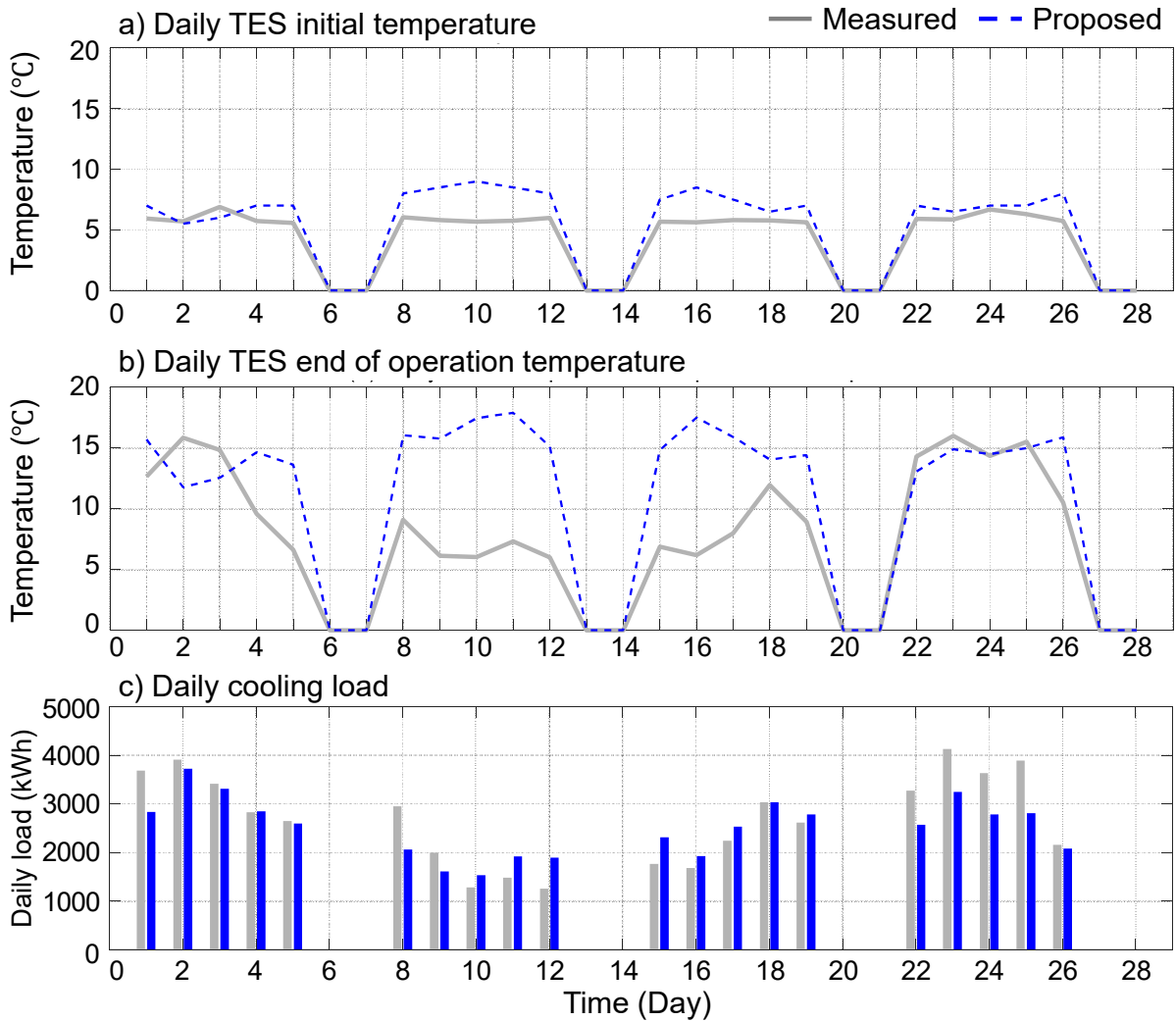

Fig. 18. Comparison of daily simulation results between conventional and proposed control strategy

# 6. Conclusion

This study developed a physics-based integrated simulation framework that can maximize the chilled water supply temperature, while ensuring cooling stability, to maximize the energy efficiency of thermal storage systems coupled with water source heat pumps. The proposed framework organically integrates four submodels: an indoor relative humidity prediction model, an RC-based load model, a PIML-based coil model, and a state-space-based TES discharge model. The dynamic changes in the chilled water limit temperature were simulated, and the optimal thermal storage setpoint temperature was derived.

The developed submodels were evaluated to obtain a suitable prediction performance for control purposes. The load model showed some errors in instantaneous prediction but demonstrated very high reliability in daily cumulative heat quantity prediction, which is a key factor in determining the TES operating capacity. Additionally, the TES model verified the control reliability by precisely simulating the thermal behavior at the peak load time when the system stability was threatened and at the discharge end time when the stratification collapsed.

Based on these validated models, a demand-side potential analysis confirmed that conventional fixed-temperature operating methods set excessive safety margins under partial-load conditions. Theoretically supply temperature increases of approximately 4.52 °C or more were possible, but when comprehensively considering physical discharge performance constraints of the TES, approximately 2.55 °C of increased operation was analyzed to be practically feasible, presenting the possibility of minimizing system surplus margin and maximizing energy utilization efficiency.

In particular, the framework proposed in this study is differentiated in that it can determine optimal control target values before the operation starts (day-ahead) by simulating the physical behavior in advance based on forecasted weather data and building information, rather than reactive response methods dependent on actual operating results. This provides a methodological foundation for eliminating uncertainty in advance through predictive control, which reflects building dynamic characteristics and establishes heat source plans that minimize safety margins.

The proposed decision-making framework verifies the effectiveness of a model-based optimal control approach capable of deriving operating strategies that adapt to load changes, unlike conventional static schedule-based operating methods. Simulation results utilizing actual operational data show that the proposed method has significant energy-saving potential, suggesting its applicability to actual building energy management systems.

Nevertheless, this study has several limitations. First, the simulation was performed under the assumption that the entire building load was proportional to the single reference floor load pattern, which failed to fully reflect the floor-specific imbalances or local load variation characteristics. Additionally, this study focused on developing an integrated simulation framework and methodological validity verification to identify the physical upper limit of the feasible chilled water supply temperature. Specific electricity consumption calculation models based on actual heat pump compressor behavior or partial load characteristics were excluded from the research scope.

Furthermore, the prediction range of data-driven models has inherent limitations. Because the proposed model learned the majority of patterns where stratification was maintained, limitations arose in predicting the stratification collapse and nonlinear temperature behavior under high-load conditions exceeding the effective thermal capacity. From a validation perspective, performance verification using the proposed integrated framework was limited to simulation environments describing historical data; on-site validation through actual field control system implementation was not performed.

Future research plans include applying the developed framework to actual building control systems to demonstrate control stability and energy-saving performance in the field. Based on the control algorithm derived from the proposed framework, the partial load operating data before and after application were comparatively analyzed to quantitatively evaluate the COP changes of the heat pump and assess the system's practical feasibility.

**Nomenclature**

| | | *Greek Symbols* | |
|---|---|---|---|
| A | Area (m$^2$) | $\alpha$ | Heat transfer coefficient (W/m$^2$·K) |
| C | Thermal capacitance (J/K) | $\varepsilon$ | Heat exchanger effectiveness (−) |
| cp | Specific heat capacity (kJ/kg·K) | $\rho$ | Density [kg/m$^3$] |
| e | Vapor pressure (Pa) | Subscripts | |
| es | Saturation vapor pressure (Pa) | a | Air |
| LMTD | Log mean temperature difference (°C) | AHU | Air handling unit |
| ṁ | Mass flow rate (kg/h) | coil | Cooling coil |
| NTU | Number of transfer units (−) | e | Envelope |
| Q | Heat transfer rate (kW) | in | Inlet |
| R | Thermal resistance (K/W) | int | Internal heat gain |
| RH | Relative humidity (%) | max | Maximum |
| T | Temperature (°C) | meas | Measured |
| UA | Overall heat transfer coefficient (kW/K) | OA | Outdoor air |
| V | Volume (m$^3$) | out | Outlet |
| | | pred | Predicted |
| | | r | Room |
| | | RA | Return air |
| | | ref | Reference |
| | | SA | Supply air |
| | | set | Setpoint |
| | | sol | Solar |
| | | TES | Thermal energy storage |
| | | w | Water |

**Declaration of Competing Interest**

The authors declare that they have no competing financial interests or personal relationships that may have influenced the work reported in this study.

**Acknowledgements**

This work was supported by an INHA UNIVERSITY Research Grant.

**Appendix**

The Appendix analyzes the frequency and load magnitude of stratification collapse inside the TES during the analysis period and the temperature distribution when the stratification collapses versus when it is maintained.

Fig. A.1 presents the time-series results of the daily cumulative cooling load and stratification collapse status at the end of operation throughout the entire simulation period. Here, stratification collapse was defined as the state where the temperature difference between the top and bottom of the TES decreased to 1°C at end of operation, resulting in loss of the thermal separation layer. The analysis results showed that stratification collapse occurred on a total of 9 out of 62 days (approximately 14.5%), and these were mostly concentrated on high-load days, falling within the top 10%.

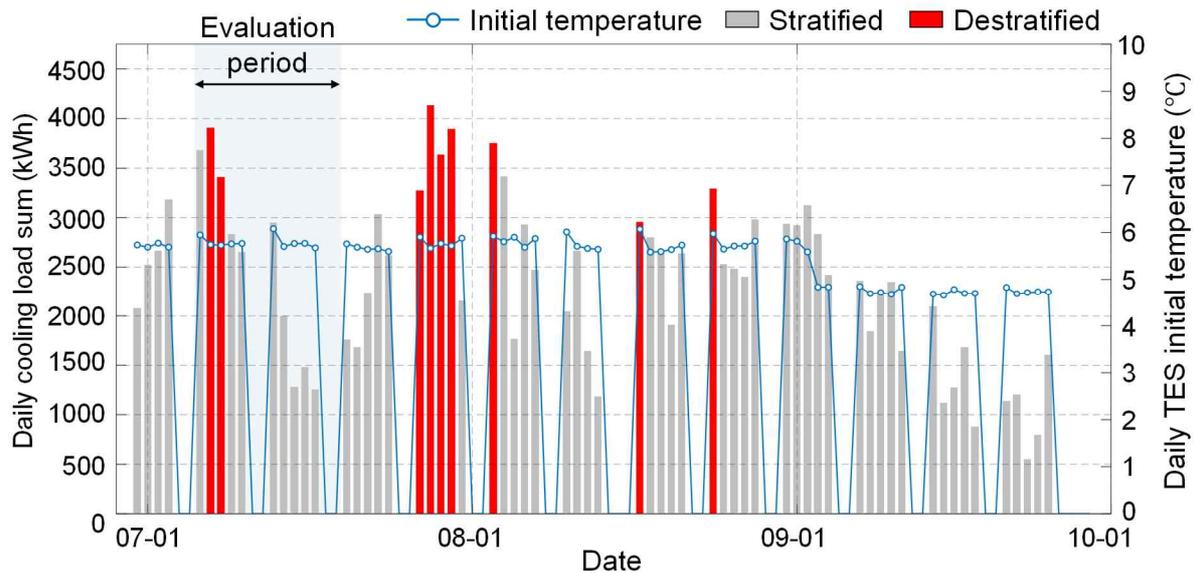

Fig. A.1 Daily cooling load profiles and stratification: simulation period representing Fig. 18

Fig. A.2 presents the comparative analysis results of the vertical temperature gradients inside the TES at the end of operation between the days when stratification was maintained and when it collapsed. When stratification was maintained, the lower part of the TES maintained approximately 5–7°C and the upper part maintained approximately 8–9°C, confirming formation of a distinct temperature boundary layer (thermocline). This indicates that the internal flow in the TES was maintained with stable stratified flow characteristics without turbulent diffusion during the operation.

However, when stratification collapses, it is interpreted as a result of accelerated fluid mixing inside the TES owing to the instability induced by the excessive load response. Accordingly, a uniform high-temperature distribution ranging from approximately 14 to 16°C from top to bottom appeared, showing typical complete mixing characteristics. In conclusion, because the proposed model was trained based on stratification maintenance patterns that constitute the majority (approximately 85.5%) of the total data, it showed a tendency to predict lower layer temperatures than the actual ones by assuming that thermal stratification is maintained even in extreme situations where stratification collapsed like day 2 in Fig. 18. Therefore, it is more appropriate to interpret the deviation between the model-predicted and actual measured values occurring in that section as originating from the physical discrepancy between the stratification collapse phenomenon occurring under operating conditions exceeding the physical limit loads and the idealized stratification behavior assumed by the model rather than as a performance defect of the model itself.

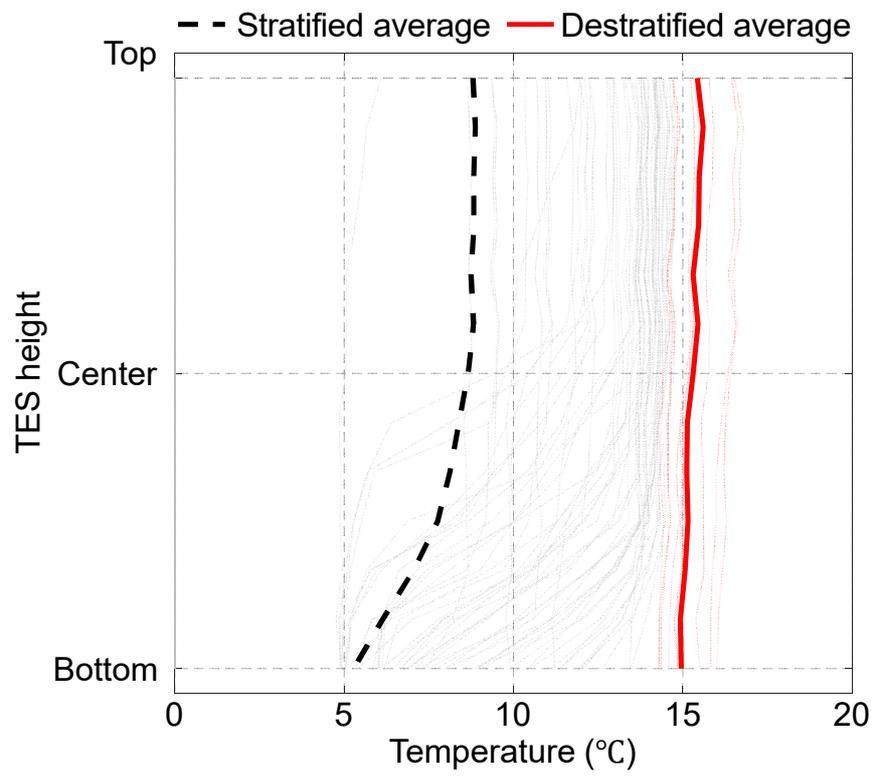

Fig. A.2 Vertical temperature profile at the end of operation